\pdfoutput=1
\documentclass[aps,prd,twocolumn,superscriptaddress,preprintnumbers,floatfix,nofootinbib]{revtex4-1}

\usepackage{graphicx}
\usepackage{amsmath}
\usepackage[caption=false]{subfig}
\usepackage{siunitx}
\usepackage{placeins}
\usepackage{color}
\usepackage{standalone}
\usepackage{dcolumn}
\usepackage{tensor}
\usepackage{bm}
\usepackage{microtype}
\usepackage{etoolbox}
\usepackage{amssymb}
\usepackage{mathrsfs}
\usepackage{accents}
\usepackage[normalem]{ulem}
\usepackage[dvipsnames]{xcolor}
\usepackage[colorlinks,urlcolor=NavyBlue,citecolor=NavyBlue,linkcolor=NavyBlue,pdfusetitle]{hyperref}
\usepackage[inline]{enumitem}

\usepackage{tikz}
\usetikzlibrary{arrows.meta}
\ifdefined\myext
\usetikzlibrary{external}
\fi

\newcommand{\beq}{\begin{equation}}
\newcommand{\eeq}{\end{equation}}

\newcommand{\infd}{{\rm d}}
\newcommand{\tgw}{T}
\newcommand{\rc}{{\rho_{\rm critical}}}
\newcommand{\rgw}{{\rho_{\rm GW}}}
\newcommand{\ogw}{{\Omega_{\rm GW}}}
\newcommand{\dir}{\hat{n}}
\newcommand{\cs}{{\rm CS}}
\newcommand{\hf}{{\tilde{h}}}
\newcommand{\normk}{\lvert\vec{k}\rvert}
\newcommand{\rlpre}{\varepsilon}
\newcommand{\fvec}[1]{{\bf #1}}
\newcommand{\lp}{A}
\newcommand{\cp}{\hat{A}}
\newcommand{\pont}{{}^{*}\!RR}
\newcommand{\obd}{\omega_{\rm BD}}
\newcommand{\dk}{\widetilde{\infd \fvec{k}}}

\newcommand{\nn}{\nonumber}

\newcommand{\mG}{\mu}
\newcommand{\htr}{\bar{h}}

\newcommand{\ifo}{I}

\newcommand*{\eq}[1]{Eq.~\eqref{eq:#1}}

\newcommand*{\sect}[1]{Sec.~\ref{sec:#1}}

\newcommand{\act}{{\cal S}}
\newcommand{\PSD}{S}
\newcommand{\PSDsky}{\PSD}

\newcommand{\Stuck}{{St\"uckelberg}}

\makeatletter
\newsavebox{\@brx}
\newcommand{\llangle}[1][]{\savebox{\@brx}{\(\m@th{#1\langle}\)}%
  \mathopen{\copy\@brx\kern-0.5\wd\@brx\usebox{\@brx}}}
\newcommand{\rrangle}[1][]{\savebox{\@brx}{\(\m@th{#1\rangle}\)}%
  \mathclose{\copy\@brx\kern-0.5\wd\@brx\usebox{\@brx}}}
\makeatother

\newtoggle{commentsoff}

\ifdefined\nocomments
    \toggletrue{commentsoff}
\fi

\iftoggle{commentsoff}{
  \newcommand*{\leo}[1]{}
  \newcommand*{\mi}[1]{}
  \newcommand*{\comment}[1]{}
  \newcommand*{\warn}[1]{}

}{
  \newcommand*{\leo}[1]{{\color{magenta} [{\bf LEO}: #1]}}
  \newcommand*{\mi}[1]{{\color{RedOrange} [{\bf MAX}: #1]}}
  \newcommand*{\comment}[1]{{\color{blue} [{\bf NOTE}: #1]}}
  \newcommand*{\warn}[1]{{\color{red} [{\bf WARNING}: #1]}}

}

\newcommand{\dcc}{LIGO-P1700234}

\begin{document}

\preprint{\dcc}

\title{Measuring stochastic gravitational-wave energy beyond general relativity}

\author{Maximiliano Isi}
\email[]{misi@ligo.caltech.edu}
\affiliation{LIGO Laboratory, California Institute of Technology, Pasadena,
California 91125, USA}

\author{Leo C.\ Stein}
\email[]{leostein@tapir.caltech.edu}
\affiliation{Theoretical Astrophysics, Walter Burke Institute for Theoretical
Physics, California Institute of Technology, Pasadena, California 91125, USA}

\hypersetup{pdfauthor={Isi and Stein}}

\date{\today}

\begin{abstract}
Gravity theories beyond general relativity (GR) can change the properties of gravitational waves:
their polarizations, dispersion, speed, and, importantly, energy content are all heavily theory-dependent.
All these corrections can potentially be probed by measuring the 
stochastic gravitational-wave background.
However, most existing treatments of this background beyond GR
overlook modifications to the energy carried by gravitational waves, or rely on GR assumptions that
are invalid in other theories.
This may lead to mistranslation between the observable cross-correlation of detector outputs and gravitational-wave
energy density, and thus to errors when deriving observational
constraints on theories.
In this article, we lay out a generic formalism for stochastic gravitational-wave
searches, applicable to a large family of theories beyond GR.
We explicitly state the (often tacit) assumptions that go into
these searches, evaluating their generic applicability, or lack
thereof.
Examples of problematic assumptions are: statistical independence of
linear polarization amplitudes; which polarizations satisfy
equipartition; and which polarizations have well-defined phase
velocities.
We also show how to correctly infer the value of the stochastic energy density in the context of any given theory.
We demonstrate with specific theories in which some of the traditional
assumptions break down: Chern-Simons gravity, scalar-tensor theory,
and Fierz-Pauli massive gravity.  In each theory, we show how to
properly include the beyond-GR corrections, and how to interpret observational results.
\end{abstract}

\maketitle

\section{Introduction}

Besides transient signals, like those detected so far \cite{gw150914, gw151226, o1bbh, gw170104, gw170608, gw170814, gw170817} by the Advanced Laser Interferometer Gravitational-wave Observatory (aLIGO) \cite{TheLIGOScientific:2014jea} and Virgo \cite{TheVirgo:2014hva}, gravitational-wave (GW) detectors are also expected to be sensitive to a persistent stochastic background \cite{Grishchuk1976,Michelson1987, Christensen1990, Christensen1992, Flanagan1993, Allen1996a, Allen1999, stochastic2017}.
This background signal is expected from primordial cosmological processes \cite{Starobinsky1979, Easther:2006vd, Barnaby:2011qe,  Cook:2011hg, Turner:1996ck, Easther:2006gt, Kamionkowski:1993fg, Gasperini:1992em, Gasperini:1993hu, Gasperini:2016gre,Caprini2018}, or the incoherent addition of myriad individually-unresolvable astrophysical sources, like compact binary coalescences \cite{Zhu:2012xw, Marassi:2011si, Lasky:2013jfa, Rosado:2012bk, Zhu:2011pt, Marassi:2009ib, Buonanno:2004tp, Sandick:2006sm} or exotic topological defects \cite{Kibble:1976sj, Damour:2004kw, Siemens:2006yp, Sarangi:2002yt}.
Among many other rich scientific goals (see \cite{Maggiore2000a} for a review)
detection of a stochastic background would provide an invaluable opportunity to study the fundamental nature of gravitational waves as they propagate over cosmological distances.

In the past decade or so, the formalism underlying stochastic GW searches has been extended to theories of gravity beyond general relativity (GR), primarily to account for the potential presence of nontensorial polarizations.
 Generic metric theories of gravity allow for up to six polarizations, corresponding to scalar (helicity 0), vector (helicity $\pm 1$) and tensor (helicity $\pm 2$) metric perturbations \cite{Eardley1973a, Eardley1973b}.
The effect of these extra polarizations on the stochastic background has been studied in particular for theories with scalar modes \cite{Maggiore2000, Gasperini2001, Nishizawa2013}, and in general for all possible modes in a theory-agnostic way \cite{Nishizawa2009, Nishizawa2010}.
The problem of detecting nontensorial modes in the background has been studied in the context pulsar timing \cite{Lee2008,Chamberlin2012,Gair2015,Cornish:2017oic}, and GW measurements using astrometry \cite{OBeirne:2018slh}.
Beyond these proposals, a comprehensive data analysis framework has been recently implemented to search LIGO and Virgo data for GWs of {\em any} polarization, tensorial or otherwise, and some first upper limits have been placed on their amplitudes \cite{Callister2017, nongrsgwo1}.

The goal of searches for stochastic backgrounds, within GR or beyond, is to measure the amount of energy that the Universe contains in the form of gravitational waves.
Consequently, treatments of stochastic GW signals are predominantly parametrized in terms of their {\em effective energy-density spectrum} [$\ogw$, defined in \eq{ogw} below].
Such parametrization is only possible thanks to a standard set of assumptions about the properties of gravitational waves, the detectors, and the statistics of the background itself.
Although generally justified within GR, the fundamental structure of beyond-GR theories may not always warrant all (or any) of those standard assumptions---even without considering modifications to specific emission mechanisms, or expected source populations.
One must therefore be careful in applying the usual premises to searches for stochastic waves that aim to be theory agnostic, and should be aware that adopting any of these assumptions may come with additional observational restrictions.

Perhaps the most important example of an assumption that has been dubiously applied beyond GR concerns the form of the effective stress-energy of GWs. 
Multiple studies of stochastic signals beyond GR assume that the fractional energy density spectrum in GWs is related to the wave amplitudes in the same way as it is in GR \cite{Nishizawa2009, Nishizawa2010, Nishizawa2013, Chamberlin2012, Yunes2013,Cornish:2017oic,OBeirne:2018slh}.
Yet, as pointed out in \cite{Stein2011}, the expression for the effective GW stress-energy need not be the same in all theories of gravity.
This means that it is inadvisable to parametrize putatively model-independent searches for beyond-GR backgrounds assuming the GW energy density has the same functional form as in GR: doing so will result in the use of a quantity that should {\em not} generally be interpreted as the energy density in GWs.
This is not only misleading, but (most importantly) can lead to incorrect comparisons between observational limits and theoretical predictions. 

Besides this, some of the simplifying assumptions about the properties of the stochastic background that are usually justified in GR are not acceptable in general, and should not be extended to model-independent analyses.
This is the case even without considering changes to the potential sources of the background in beyond-GR theories, which may themselves break more of the assumed symmetries.
For instance, it is not reasonable to always assume that the usual linear GW polarization amplitudes will be statistically independent, as this will not be true unless the chosen polarization basis diagonalizes the kinetic matrix of the underlying theory of gravity.
Similar arguments can be made about the assumptions that the polarizations are equipartitioned, or even that they have well defined phase velocities---let alone that they propagate at the speed of light. 

In view of this, our goal is to straighten out the framework underlying searches for stochastic gravitational backgrounds, to make it formally valid and easily applicable to a large family of theories beyond GR.
In \sect{formalism}, we lay out a generic formalism for such searches, review the most commonplace assumptions in standard analyses, and evaluate their degree of applicability to other frameworks;
along the way, we also clarify some relevant differences in conventions used by the theory and data analysis literatures.
In \sect{examples}, we provide a series of examples of theories that break the premises behind one or more of these assumptions, and show the impact this has on the analysis---we focus on differences in the predicted form of the effective GW stress energy, but also discuss other problematic points.
In particular, we use these examples to show how to go from the action defining a theory to {\em (1)} a relation between the fractional GW energy density spectrum and the correlation of polarization amplitudes, and {\em (2)} to the cross-correlation of GW detector outputs---which is the relevant observable for ground-based detectors. 
We review the derivation for general relativity in \sect{gr}, and then
move on to Chern-Simons gravity in \sect{cs}, scalar-tensor theories
in \sect{st}, and Fierz-Pauli massive gravity in \sect{mg}.
Finally, we offer a summary and conclusions in \sect{conclusion}.

\section{Formalism}
\label{sec:formalism}

In this section, we provide the framework required to search for stochastic GW backgrounds without assuming GR is correct.
In \sect{decomp}, we review the four-dimensional Fourier transform of a generic GW, lay out its decomposition into polarizations, and provide some useful identities for later use in \sect{examples}.
In \sect{stochastic}, we focus on the properties of stochastic backgrounds, carefully reviewing the assumptions made in traditional analyses to determine whether they hold in theories beyond GR.
In \sect{detection}, we describe the measurement process, including complications that may arise in generic theories.
Finally, in \sect{energy}, we sketch the calculations needed to relate the effective stochastic GW energy in any given theory to the polarization amplitudes measurable by a detector.

Here, and throughout this document, spatial three-vectors are identified by an arrow (e.g.\ $\vec{k}$), or a circumflex accent if they have unit norm (e.g.\ $\hat{k}$).
Four-vectors and higher-rank tensors are denoted by boldface, or abstract index notation (e.g.~$\fvec{k}$ or $k_a$).
For tensor coordinate components, spacetime Greek indices ($\alpha$, $\beta$, $\gamma$, \ldots{}) take values in the range 0--3, while spatial Latin indices ($i$, $j$, $k$, \ldots{}) span 1--3.
We use metric signature $+2$, using $g_{ab}$ for generic background metrics and $\eta_{ab}$ for the Minkowski metric. 
Our conventions for the Levi-Civita tensor follow~\cite{MTW}: $\epsilon_{abcd} = \sqrt{-g}[abcd]$ where $g$ is the determinant of the metric, and $[abcd]$ is the Levi-Civita \emph{symbol}, with $[0123] = +1$; similarly, $\epsilon_{ijk} = \sqrt{\gamma}[ijk]$ where $\gamma$ is the determinant of the spatial metric, and $[123]=+1$.
We normalize \mbox{(anti-)symmetrizations} as idempotent projection
operations, e.g., $T_{(ab)} = \left(T_{ab} + T_{ba}\right)/2$ and
$T_{[ab]} = \left(T_{ab} - T_{ba}\right)/2$.

\subsection{Decomposition of the metric perturbation} \label{sec:decomp}

In any metric theory of gravity, as long as the observation region is small compared to the curvature radius, an arbitrary GW metric perturbation $h_{ab}(\fvec{x})$ at a spacetime point $\fvec{x}$ may be expressed as a plane-wave expansion by the compact expression:
\beq \label{eq:h_expansion}
h_{ab}(\fvec{x}) = \frac{1}{2\pi} \int \hf_{ab}(\fvec{k}) e^{i\fvec{k} \cdot
\fvec{x}}~\dk\, ,
\eeq
integrating over all directions of propagation, and over both positive and negative frequencies.
Here $\hf_{ab}(\fvec{k})$ is the complex-valued Fourier amplitude for the wave-vector $\fvec{k} \equiv (\omega/c,\,\vec{k})$;
we let $\omega=2\pi f$ be the angular frequency, and $\vec{k}=\normk\, \hat{k} \equiv -\normk\, \dir$ the spatial wave-vector, implicitly defining $\dir$ as the sky location of the source.
To simplify our notation in \eq{h_expansion}, we have defined the four-dimensional integral over the measure
\beq \label{eq:dk}
\dk \equiv 2 c\, \delta(|\vec{k}|^2 - |\vec{k}_\omega|^2)\, |\vec{k}|^{-1} \infd
\fvec{k} = \infd \omega\, \infd \dir\, ,
\eeq
where $\delta(x)$ is the Dirac delta function, and the last equality assumes an implicit integration over the magnitude of $\vec{k}$.
In order to write this, we assume that there is just {\em one} dispersion relation, $\omega=\omega(\vec{k})\equiv \omega_k$, that determines the modulus of $\fvec{k}$ and implicitly defines $|\vec{k}_\omega| \equiv |\vec{k}|(\omega)$.
\footnote{Some theories violate this assumption; for example, bimetric gravity~\cite{Hassan:2011zd} has one massless and one massive gravitational wave mode---we will allow for this briefly in \sect{detection} only.}
The dispersion relation is specific to the theory of gravity: for example, $\omega_k = c\normk$ and $|\vec{k}_\omega| = \omega/ c$ in GR.

With the integration measure defined as in \eq{dk}, in a local Lorentz frame (so that $\fvec{x}\cdot\fvec{y} = \vec{x}\cdot\vec{y}-x_0y_0$), \eq{h_expansion} can be recast in a form most common in stochastic GW literature (see, e.g., \cite{Flanagan1993, Allen1999, Romano2017}):
\beq \label{eq:h_expansion_long}
h_{ab}(t, \vec{x}) = \int_{-\infty}^{\infty} \int_{\rm sky} \hf_{ab}(f, \dir)\,
e^{-2\pi i f(t + \dir \cdot\vec{x}/v_{\rm p})}\, \infd \dir\, \infd f\, ,
\eeq
where $v_{\rm p} \equiv |\vec{k} / \omega|^{-1}$ is the (potentially frequency-dependent) phase velocity of the wave ($v_{\rm p}=c$ in GR).
Finally, to guarantee that $h_{ab}(\fvec{x})$ be real, we must necessarily have
\beq \label{eq:ftconj}
\hf^*_{ab}(f,\dir) = \hf_{ab}(-f, \dir)\, ,
\eeq
where the asterisk indicates complex conjugation.
In Appendix \ref{app:planewave}, we elucidate the equivalence between Eqs.\ \eqref{eq:h_expansion} and \eqref{eq:h_expansion_long}, derive the second equality in \eq{dk} and discuss differences between our Fourier conventions and those from the field theory literature.

For any given frequency and direction of propagation, the Fourier amplitudes may be written as a linear combination of at most six tensors corresponding to the six polarizations supported by generic metric theories of gravity~\cite{Eardley1973a, Eardley1973b}, even if the wave speed is slightly different from the speed of light~\cite{Will:1993ns}.
Therefore, the most generic gravitational wave in this large category of theories may be written as a function of six independent amplitudes.
We may, therefore, define six orthogonal polarization tensors,  $e^{\lp}_{ab}$, such that
 \beq \label{eq:poldecomp}
\tilde{h}_{ab} (\fvec{k}) = \tilde{h}_{\lp}(\fvec{k})\, e^{\lp}_{ab}(\dir)~,
\eeq
where the sum is over six polarizations indexed by $\lp$, and the $\hf_{\lp}(\fvec{k})$'s are the Fourier transforms of the six scalar fields, $h_{\lp}(\fvec{x})$, encoding the amplitude of each mode, as defined by means of \eq{h_expansion}

In order to study interactions between waves and detectors, it is usually convenient to pick a ``synchronous'' gauge%
\footnote{In a diffeomorphism invariant theory, one may always gauge transform into synchronous gauge by solving an initial value problem. If the theory is not diff-invariant, the \Stuck{} trick can be used to restore the symmetry and then gauge transform. We provide an example of this in \sect{mg}.}
such that the perturbation is purely spatial in the frame of interest ($h_{0\nu}=0$), and correspondingly so are the polarization tensors.
For instance, in an orthogonal frame in which the $z$-axis is aligned with the direction of propagation (so that $\hat{k}_i = \delta^3_{~i}$ in that frame), we may write the six degrees of freedom as
\beq \label{eq:polarizations}
(h_{ij}) = \begin{pmatrix}
h_{\rm b} + h_+ & h_\times    & h_{\rm x}  \\
h_\times    & h_{\rm b} - h_+ & h_{\rm y}  \\
h_{\rm x}    & h_{\rm y}    & h_{\rm l}
\end{pmatrix} ,
\eeq
in terms of the linear tensor polarizations ($+$, $\times$), linear vector polarizations (x, y), and scalar breathing (b) and longitudinal (l) modes.

For the purpose of analyzing the output of multiple GW detectors, it
is often convenient to write the polarization tensors in terms of unit
vectors tangent and normal to the celestial sphere at each sky location.
A standard linear polarization basis is given by
\begin{subequations} \label{eq:poltensors}
  \begin{align}
\label{eq:plus}
e^+_{ab}(\dir) &=  \hat{\phi}_a(\dir)\, \hat{\phi}_b (\dir) -
\hat{\theta}_a(\dir)\, \hat{\theta}_b(\dir)\, ,
\\
\label{eq:cross}
e^\times_{ab}(\dir) &= \hat{\phi}_a(\dir)\,  \hat{\theta}_b(\dir)+ 
\hat{\theta}_a(\dir)\, \hat{\phi}_b(\dir)\, ,
\\
\label{eq:vectorx}
e^{\rm x}_{ab}(\dir) &= \hat{\phi}_a(\dir)\, \hat{k}_b(\dir) + 
\hat{k}_a(\dir)\, \hat{\phi}_b (\dir)\, ,
\\
\label{eq:vectory}
e^{\rm y}_{ab}(\dir) &= \hat{\theta}_a(\dir)\, \hat{k}_b(\dir) + 
\hat{k}_a(\dir)\, \hat{\theta}_b (\dir)\, ,
\\
\label{eq:breathing}
e^{\rm b}_{ab}(\dir) &= \hat{\phi}_a(\dir)\, \hat{\phi}_b(\dir) + 
\hat{\theta}_a(\dir)\, \hat{\theta}_b (\dir)\, ,
\\
\label{eq:longitudinal}
e^{\rm l}_{ab}(\dir) &=  \hat{k}_b(\dir)\,  \hat{k}_b(\dir)\, ,
  \end{align}
\end{subequations}
where $\hat{\theta}(\dir)$ and $\hat{\phi}(\dir)$ are respectively the celestial polar and azimuthal coordinate vectors for a given source sky location determined by $\dir$;
by design, these vectors satisfy $\hat{\theta}(\dir) \times \hat{\phi}(\dir) = - \hat{k}(\dir) = \dir$.
Other frame choices are possible, and multiple conventions abound in the literature.

As an example of the polarization decomposition of \eq{poldecomp}, consider theories in which gravitational perturbations carry spin-weight 2, like GR.
In that case, we may choose to work with the two transverse-traceless linear polarization tensors corresponding to the plus ($+$) and cross ($\times$) amplitudes shown in \eq{polarizations}, and \eq{poldecomp} becomes simply:
\beq \label{eq:h_linear}
\hf_{ab}(\fvec{k}) = \hf_+(\fvec{k})\thinspace e^+_{ab}(\dir) + \hf_\times
(\fvec{k})\thinspace e^\times_{ab}(\dir)\, .
\eeq
Because the linear polarization tensors are real-valued by definition [cf.~\eq{poltensors}], the reality condition for the amplitudes, \eq{ftconj}, implies 
\beq \label{eq:reality_linear}
\hf_{+/\times}(-f, \dir)=\hf^*_{+/\times}(f,\dir)\, .
\eeq

Alternatively, instead of the linear modes of \eq{polarizations}, we could choose to work with eigenmodes of the helicity operator, i.e.\ the right- and left-handed circular polarization tensors (denoted ``R'' and ``L'' respectively).
These modes satisfy an eigenvalue equation
\begin{equation}
\label{eq:id2}
\epsilon^{ijk}\hat{k}_k e^{\cp}{}_{\ell j} =i\,\rlpre_{\cp}\,
e^{\cp}{}_\ell{}^i\, ,
\end{equation}
for $\cp\in\{\rm R, L\}$ (not summed on the RHS), where we have defined the factor $\rlpre_{\rm R/L} = \pm 1$, with the plus (minus) sign corresponding to the R (L) mode.
Then, the circular polarization tensors can be written in terms of the ones for plus and cross as
\beq \label{eq:eRL}
{\bf e}_{\rm R/L} =\frac{1}{\sqrt{2}} \left({\bf e}_+ + i\,\rlpre_{\rm R/L} {\bf e}_\times
\right) .
\eeq
Using the circular tensors as a basis, we would write, instead of \eq{h_linear},
\beq \label{eq:circdecomp}
\hf_{ab}(\fvec{k}) = \hf_{\rm R}(\fvec{k})\, e^{\rm R}_{ab} (\dir) +
\hf_{\rm L}(\fvec{k}) \, e^{\rm L}_{ab} (\dir)\, ,
\eeq
and the reality condition, \eq{ftconj}, would now imply (note the ``L/R'' subscript on the right hand side)
\beq \label{eq:reality_circular}
\hf_{\rm R/L}(-f, \dir) = \hf^*_{\rm L/R}(f,\dir)\, ,
\eeq
instead of \eq{reality_linear}.
The circular polarization modes can be similarly defined for the vector polarizations to obtain eigenmodes of helicity $\pm1$.
On the other hand, the scalar modes have helicity 0, so in a sense are already circular.

For future reference, note that the spin-weight 2, spin-weight 1 and the transverse spin-weight 0 linear polarization tensors are normalized as usual such that, for a given direction of propagation,
\beq \label{eq:lp_norm}
e^{\lp \, ij} e^{\lp'}{}_{ij}= 2\, \delta^{\lp\lp'}\, ,
\eeq
for $\lp \in \{+,\,\times,\,{\rm x},\,{\rm y},\,{\rm b}\}$, and $\delta^{\lp\lp'}$ the 
Kronecker delta; on the other hand, the longitudinal tensor satisfies $(e^{{\rm l}})^{ij} (e^{\lp})_{ij}=\delta^{{\rm l}\lp}$.
Similarly, the spin-weight 2 circular polarization tensors of \eq{eRL} satisfy
\begin{align}
\label{eq:lc_norm}
(e^{\cp \, ij})^* e^{\cp'}{}_{ij} &= 2\, \delta^{\cp\cp'}\, .
\end{align}
The basis tensors for the circularly-polarized vector modes also satisfy \eq{lc_norm}.

Although the two linear and circular bases discussed above are probably the most common in the GW literature (modulo normalizations), we are of course free to pick any other.
For instance, in the analysis of differential-arm instruments, it is generally convenient to instead work with the traceless linear combination of $h_{\rm b}$ and $h_{\rm l}$, since that is what such detectors can measure.
Similarly, different theories may also define their own preferred polarization bases, given by the choice that diagonalizes their kinetic matrices.

\subsection{Stochastic signals} \label{sec:stochastic}

\newcounter{AssumptCounter}
\renewcommand{\theAssumptCounter}{\roman{AssumptCounter}}
\newcommand*{\alabel}[1]{({\em \refstepcounter{AssumptCounter}\theAssumptCounter\label{a:#1}})}
\newcommand*{\aref}[1]{({\em \ref{a:#1}})}

In the case of stochastic signals, the Fourier amplitudes, $\hf_{ij}(\fvec{k})$, are, by definition, random variables and, as such, can be fully characterized by the moments of some (multivariate) probability distribution.
Most standard searches for a stochastic GW background make the following assumptions about the random process that produced these amplitudes (see, e.g., \cite{Romano2017} for a review):
the random process is
\aref{gaussian} Gaussian,
\aref{ergodic} ergodic,
and
\aref{stationary} stationary,
with no correlation between amplitudes from different 
\aref{indepsky} sky locations or 
\aref{indeppol} polarizations, and with 
\aref{equipartition} equipartition of power across polarizations; furthermore, the process is commonly (although not universally) assumed to be 
\aref{isotropic} isotropic.
We break down these assumptions below, and introduce some important definitions along the way.

Stochastic backgrounds are expected to arise from primordial cosmological processes \cite{Starobinsky1979, Easther:2006vd, Barnaby:2011qe,  Cook:2011hg, Turner:1996ck, Easther:2006gt, Kamionkowski:1993fg, Gasperini:1992em, Gasperini:1993hu, Gasperini:2016gre,Caprini2018}, or by the incoherent superposition of a great number of signals from contemporary astrophysical events \cite{Zhu:2012xw, Marassi:2011si, Lasky:2013jfa, Rosado:2012bk, Zhu:2011pt, Marassi:2009ib, Buonanno:2004tp, Sandick:2006sm, Kibble:1976sj, Damour:2004kw, Siemens:2006yp, Sarangi:2002yt,Maggiore2000a}.
The assumption \alabel{gaussian} that the astrophysical background is produced by a Gaussian random process is motivated by the central limit theorem---this guarantees that the properties of any large number of incoherently-added GW signals will be normally distributed, regardless of the specific characteristics of any given source.
A similar argument can be applied to primordial signals by considering the independent evolution of waves from causally-disconnected regions \cite{Allen1996a}.
Although waves from inflation will technically have non-Gaussianities, they will be small as long as inflation satisfied the slow-roll approximation~\cite{Maldacena2002,Caprini2018,Allen1996a}.

For Gaussian processes, all properties of the probability distribution are determined by its first two moments (correlation functions)---namely, the mean and power spectrum (respectively, the one- and two-point correlation functions).
The first moment of the distribution, the mean $\langle \hf (f) \rangle$, will not appear explicitly in any of the expressions below, so we ignore it.%
\footnote{Some authors explicitly set this value to zero because the contribution from a nonvanishing mean would take the form of a {\em coherent} offset in the Fourier amplitudes as a function of frequency, which not only would be hard to justify physically, but would also hardly classify as ``stochastic'' (see, e.g., \cite{Romano2017,Allen1999}).}
Here and below, the expectation value, denoted by angle brackets $\langle \cdot \rangle$, corresponds to ensemble averages, as well as space/time-averages by assumption \alabel{ergodic} of ergodicity.
The expectation of ergodicity itself comes from the assumption that the Universe is homogeneous (for more discussion on this topic, see \cite{Caprini2018}).

The second moment of the distribution will end up being an important observable.
In order to write down an expression for it, we can make use of assumptions \aref{stationary} and \aref{indepsky}.
First, stationarity \alabel{stationary} is motivated by the fact that observation times (order of months to years) are extremely small relative to the dynamical timescales intrinsic to the cosmological processes that could change the properties of the background (order of billions of years); therefore, any changes in the stochastic background would be unnoticeable to us.
Formally, stationarity means that the first moment is constant, while the second moment depends only on time {\em differences} (see, e.g., \cite{Jaranowski2009}).
As a consequence, the Fourier transform of a stationary random variable can be shown to be such that amplitudes at different frequencies will be statistically independent and, therefore, uncorrelated (Appendix \ref{app:correlation}).

Next, the assumption \alabel{indepsky} that amplitudes from different
sky locations will be uncorrelated is justified for primordial waves
because signals from different points in the sky are only coming into
causal contact now at Earth, under ordinary topological assumptions.
One could potentially search for nonstandard spatial topologies in a
sufficiently ``small'' universe through angular correlations in
gravitational waves~\cite{Grishchuk:1975ec}, in much the same way as
in the cosmic microwave background (CMB)~\cite{LachiezeRey:1995kj, Cornish:1997ab}.  A small
universe with nonstandard spatial topology would induce circles of
excess correlation in both the CMB and gravitational-wave background.
As there has been no evidence of this phenomenon in the CMB, in this
article we consider primordial signals from different sky directions
to be uncorrelated.

For contemporary (``astrophysical'') backgrounds, \aref{indepsky} comes from the assumption that the contributions from multiple sources throughout the sky (say, binary systems) are added ``incoherently''---that is, sources are not perfectly aligned and timed as would be needed for signals from different directions to reach us with matching phase and amplitude evolution.
Even though such astrophysical sources were in causal contact at some
point in the past, they are embedded in chaotic astrophysical
environments (with e.g.~turbulent magnetohydrodynamics) with Lyapunov
times sufficiently short that in practice, they can be treated as
uncorrelated.
In principle, strong gravitational lensing may introduce correlations
between sky bins into the stochastic background, whether primordial or
contemporary, but we can expect this effect to be negligible in
practice~\cite{Smith:2017mqu}.

With assumptions \aref{stationary} and \aref{indepsky} in place, we may write the second moment of the amplitude distribution in the form (Appendix \ref{app:correlation}):
\beq \label{eq:correlation}
\left\langle \hf^*_A(\fvec{k})\, \hf_{A'}(\fvec{k}') \right\rangle = \frac{1}{2} \delta(f-f')\, \delta(\dir - \dir')\, \PSDsky_{AA'}(\fvec{k})\, . 
\eeq
This equation defines the {\em one-sided cross-power spectral density}, $\PSDsky_{AA'}(\fvec{k})\equiv\PSDsky_{AA'}(f,\dir)$, for two signals, $\hf_{A/A'}(\fvec{k})$, sharing a wave-vector $\fvec{k}$ but with potentially different polarizations $A$ and $A'$.
For linear polarizations, this quantity satisfies $\PSDsky_{\lp\lp'}(f, \dir) = \PSDsky_{\lp'\lp}(-f, \dir)$, because of the reality condition of \eq{ftconj}.
Since we are usually interested in the total measured power at a given frequency, regardless of sky direction, we also define the integral of $\PSDsky_{AA'}(\fvec{k})$ over the sky,
\beq \label{eq:saa}
\PSD_{\lp\lp'}(f) \equiv \int_{\rm sky} \PSDsky_{\lp\lp'}(f,\dir) \thinspace \infd \dir~,
\eeq
which carries units of ${\rm strain}^2/{\rm Hz}$. 
For $\lp=\lp'$, this is nothing more than the one-sided {\em power spectral density} (PSD) in polarization $\lp$, which we denote $\PSD_{\lp}(f) \equiv \PSD_{\lp\lp}(f)$.
In general, for any real-valued random variable $X(t)$, the PSD can be approximated as twice the square of the band-limited Fourier transform \cite{Jaranowski2009},
\beq \label{eq:psd}
\PSD_X(f) = \lim_{T\rightarrow\infty} \frac{2}{T} \left| \int_{T/2}^{T/2} X(t)\, e^{2\pi i f t} \infd t \right|^2,
\eeq
in practice always computed for some long but finite integration time, $T$, on the order of months to years for observations of the stochastic background.
As usual, the factor of 2 in \eq{psd} accounts for the fact that this is the {\em one-sided} PSD, $S(f)\equiv S(|f|)$.

Assumption \alabel{indeppol} that the different polarizations are statistically independent may be used to discard off-diagonal terms in the cross-power spectrum, so that $S_{\lp\lp'}(f)= \delta_{\lp\lp'} S_{\lp}(f)$.
However, one must be careful with this simplification: the assumption is valid {\em if and only if} one works in a polarization basis that diagonalizes the kinetic matrix of the theory.
Importantly, as we will show with specific examples, such a basis need {\em not} be the linear polarization basis used in most GR analyses.
Even when working within GR, it is generally better, from a theoretical standpoint, to work in terms of the circular modes, as they are eigenstates of the helicity operator, and they might be produced with different intensities in the early universe \cite{Alexander2006, Seto2007, Seto2008}.

Besides assuming that the polarizations are uncorrelated, it is also common to assume that there is equipartition of power between them---assumption \alabel{equipartition} in our list above.
Under this presumption, the background is said to be unpolarized and the polarization PSDs may be written in terms of the {\em total} GW spectral density, $S(f)=\sum S_A(f)$, such that $S_{\lp} = S(f)/N$, where $N$ is the number of polarizations allowed to propagate in a given theory.
In general, this assumption is only justified if the polarizations both diagonalize the kinetic matrix {\em and} interact similarly with matter, so that they are sourced in equal amounts.
This not always the case: for example, in both massive gravity \cite{DeRham2014,Hinterbichler2011} and dynamical Chern-Simons gravity \cite{Jackiw2003, Alexander2009, Sopuerta2009}, different polarizations couple to sources with different strengths.

Finally, the simplest searches for a stochastic background also adopt assumption \alabel{isotropic} of isotropy, in which case $\PSDsky(f,\dir) = \PSD(f)/4\pi$, by \eq{saa}.
In GR, if one disregards the proper motion of the solar system, this assumption is expected to hold well for most foreseeable sources of a stochastic background detectable by existing ground-based observatories, since they are expected to originate from cosmological distances
\cite{Romano2017,Allen1996a, Allen1999}.
\footnote{The same will not necessarily be true for LISA, which will be sensitive to galactic stochastic sources, like the ``confusion noise'' from white-dwarf binaries \cite{Ruiter:2007xx}.}
For cosmological sources, isotropy is likely also a good
assumption in many beyond-GR theories;  however, isotropy should not be expected to hold
in theories with a preferred frame, which are intrinsically
anisotropic~\cite{Kostelecky:1988zi, Kostelecky:1991ak,
Kostelecky:1994rn, Kostelecky:2003fs, Kostelecky2016, Jacobson:2000xp,
Eling:2004dk, Jacobson:2008aj}.
For simplicity, the rest of this document will treat only the case of an isotropic background, but this does not affect the spirit of the results, which can be easily generalized to the anisotropic case.
For predictions of the angular power spectrum of astrophysical GR backgrounds see \cite{Cusin:2018rsq}, and for corresponding observational limits that do not assume isotropy see \cite{anisostochastic2017}.

Assuming both \aref{isotropic} an isotropic background and \aref{indeppol} uncorrelated polarizations, on top of \aref{stationary} stationarity and \aref{indepsky} uncorrelated sky bins, \eq{correlation} can be written directly in terms of the power spectral density,
\beq \label{eq:correlation_isotropic}
\left\langle \hf^*_A(\fvec{k}) \hf_{A'}(\fvec{k}') \right\rangle =
\frac{1}{8\pi} \delta(f-f') \delta(\dir - \dir') \delta_{\lp\lp'} \PSD_{\lp}(f).
\eeq
If one further assumed \aref{equipartition} equipartition, $S_{\lp}(f)$
would be replaced with $S(f)/N$, as explained above.
This is the form of the expression most common in recent literature about detection of stochastic gravitational-wave backgrounds (e.g., \cite{Romano2017}).

\subsection{Detection} \label{sec:detection}

Because the output of ground-based GW detectors is largely dominated by stochastic instrumental and environmental noise \cite{gw150914detchar, Covas2018}, it is not possible to measure the power spectrum of the polarization amplitudes, $S_{\lp}(f)$, directly with a single detector at any level of interest.
However, this quantity may be inferred by looking instead at the cross-correlation of the output of two or more instruments (see, e.g., \cite{Romano2017} for a comprehensive review of data analysis methods).

We assume that each GW detector has a purely linear response to
gravitational waves.  Therefore, in the Fourier domain, the response
of detector $\ifo$ to a plane wave $\hf_{ab}(\fvec{k})$ must be
expressible as
\begin{equation}
\label{eq:hf_plane}
\hf_{\ifo}(\fvec{k}) = \tilde{D}^{ab}_{\ifo}(\fvec{k})\, \hf_{ab}(\fvec{k})\, ,
\end{equation}
for some tensor
$\tilde{D}^{ab}_{\ifo}(\fvec{k}) \equiv\tilde{D}^{ab}_{\ifo}(f,\dir)$
representing the detector's frequency- and direction-dependent
transfer function.  This tensor encodes all relevant
information about the detector and the physics of the measurement
process~\cite{Forward1978, bluebook, Estabrook1985, Schilling:1997id,
  Rakhmanov2005a, Rakhmanov:2008is, Rakhmanov:2009zz, Essick:2017wyl}
(for considerations specific to gravity beyond GR, see
e.g.~\cite{Eardley1973b, Tobar:1999tf, Blaut2012, poisson2014, Isi2017, Gair2015}).
The
detector's output $\hf_{\ifo}(\fvec{k})$ (e.g.~the calibrated current out of a
photodiode) is a gauge-invariant observable.  However, the metric
perturbation $\hf_{ab}(\fvec{k})$ is gauge-dependent; therefore, the
detection tensor $\tilde{D}^{ab}_{\ifo}(\fvec{k})$ must also depend on
the gauge choice, so that the overall gauge dependence on the RHS of
\eq{hf_plane} exactly cancels.

Assuming a basis of polarization states $A$ that have well defined phase velocities (i.e.~they diagonalize the kinetic matrix of the theory), we may use \eq{hf_plane} to write the Fourier transform of the signal at detector $\ifo$ explicitly as a sum over polarizations and an integral over sky directions,
\beq \label{eq:hf_general}
\hf_{\ifo}(f) = \int \sum_A \tilde{F}^{\lp}_{\ifo} (f, \dir)\, \hf_{\lp}(f, \dir)\, e^{-2\pi i f \dir \cdot\vec{x}_{\ifo}/v^A_{\rm p}}\, \infd \dir\, ,
\eeq
defining the Fourier-domain response functions as the contraction between the detector and polarization tensors,
$\tilde{F}^{\lp}_{\ifo} (f, \dir) \equiv \tilde{D}^{ab}_{\ifo} (f,
\dir)\,  e^{\lp}_{ab}(\dir)$, which must also be gauge-dependent.

The time-domain analogue of \eq{hf_plane} is given by a convolution,
\beq
h_{\ifo} (t, \vec{x}_{\ifo}) = \int_{-\infty}^{\infty} D^{ab}_{\ifo}(t)\, h_{ab}(t-\tau, \vec{x}_{\ifo})\, \infd \tau\, , 
\eeq
with $\vec{x}_{\ifo}$ the location of detector $I$, and $D^{ab}_{\ifo}(t)$ its impulse response.
Since $\tilde{D}^{ab}_{\ifo}(\fvec{k})$ is gauge-dependent, the same must be true for $D^{ab}_{\ifo}(t)$.
For an ideal differential arm-length instrument, it is easiest to write down this detector tensor in a synchronous gauge ($h_{0\nu}=0$ in the detector frame), wherein the end
test masses' coordinate locations will not change~\cite{MTW}.  In
such a gauge, the resulting differential-arm detector tensor is the purely geometric
factor
\beq \label{eq:detector}
D_{ab}(t) = \frac{1}{2} \left(\hat{X}_a \hat{X}_b - \hat{Y}_a \hat{Y}_b \right),
\eeq
with $\hat{X}$ and $\hat{Y}$ spacelike unit vectors pointing along the detector arms.
For real interferometric detectors, like LIGO and Virgo, \eq{detector}
is valid {\em only} in the small-antenna limit (arm length $\ll$ GW
wavelength)~\cite{Schilling:1997id, Rakhmanov2005a, Rakhmanov:2008is,  Rakhmanov:2009zz, Essick:2017wyl}.

For any realistic detector, the tensor of \eq{detector} will vary in time due to the motion of the instrument with respect to the inertial frame of the wave (e.g.\ due to Earth's rotation, for ground-based observatories).
However, for the cases we are interested in, we can take this variation to be slow with respect to the period of the waves, so that it can be ignored if \eq{hf_general} is implemented via short-time Fourier transforms.
In this ideal ``slow-detector'' limit, we may then treat the response as time- and frequency-independent to write $\tilde{D}^{ab}_I(\fvec{k})=D^{ab}_I(t)\equiv D^{ab}_I$, and so \eq{hf_general} simplifies to
\beq \label{eq:hf}
\hf_{\ifo}(f) = \int \sum_A F^{\lp}_{\ifo} (\dir)\, \hf_{\lp}(f, \dir)\, e^{-2\pi i f \dir \cdot\vec{x}_{\ifo}/v^A_{\rm p}}\, \infd \dir\, ,
\eeq
with the frequency-independent {\em antenna patterns} defined in full analogy to our definition of $\tilde{F}^{\lp}_{\ifo}(f,\dir)$ above,
\beq \label{eq:ap}
F^{\lp}_{\ifo} (\dir) \equiv D^{ab}_{\ifo}\,  e^{\lp}_{ab}(\dir)\, .
\eeq
For details on this simplification, and nuances applicable to anisotropic backgrounds, see Sect.~IV in \cite{Allen1997}.

In the Fourier domain, the cross-correlation between the output of two  detectors may then be written in terms of the second moment of the distribution of polarization amplitudes as
\begin{align} \label{eq:crosscor}
\left\langle \hf_{\ifo}^*(f) \hf_{\ifo'}(f') \right \rangle = 
&\int \infd\dir \infd\dir' \sum_{AA'} \left\langle \hf^*_{\lp}(\fvec{k}) \hf_{\lp'}(\fvec{k}') \right\rangle \\
&\times F^{*\lp}_{\ifo} (\dir) F^{\lp'}_{\ifo'} (\dir')\, e^{i (\vec{k}_{A'}
\cdot\vec{x}_{\ifo'} - \vec{k}_{A} \cdot\vec{x}_{\ifo})}\, ,\nonumber
\end{align}
where, again, assumption \aref{ergodic} of ergodicity is tacitly implied.
If we also assume, as we will throughout this paper, that the background is \aref{stationary} stationary and \aref{isotropic} isotropic, and \aref{indepsky} that sky bins are uncorrelated, we may then use \eq{correlation_isotropic} to simplify this to
\begin{align} \label{eq:crosscor_isotropic}
\left\langle \hf_{\ifo}^*(f) \hf_{\ifo'}(f') \right \rangle &= \frac{1}{2}
\delta(f-f')  S_{\lp\lp'}(f) \Gamma^{\lp\lp'}_{~~\ifo\ifo'}(f)\, ,
\end{align}
where we have defined the {\em generalized overlap reduction function} for polarizations $\lp$, $\lp'$ and detectors $\ifo$, $\ifo'$,
\beq \label{eq:overlap}
\Gamma^{\lp\lp'}_{~~\ifo\ifo'}(f) \equiv \frac{1}{4\pi} \int \infd\dir\,
F^{*\lp}_{\ifo} (\dir) F^{\lp'}_{\ifo'} (\dir) e^{-2\pi i f \dir \cdot \xi^{AA'}_{\ifo\ifo'}},
\eeq
in terms of the phase factor $\xi_{\ifo\ifo'}^{AA'}(f) \equiv \vec{x}_{\ifo}/v_{\rm p}^{A} - \vec{x}_{\ifo'}/v_{\rm p}^{A'}$, which acquires a potential frequency dependence through the phase velocities.
If there is one dispersion relation shared by all polarizations (true throughout the rest of this paper), the exponent in \eq{overlap} can be written as $-2\pi if\dir \cdot \xi_{II'}^{AA'} = -i\vec{k}\cdot \Delta \vec{x}_{\ifo\ifo'}$, in terms of the separation between detectors $\Delta \vec{x}_{\ifo\ifo'} \equiv \vec{x}_{\ifo} - \vec{x}_{\ifo'}$.
The overlap reduction functions encode all relevant information pertaining GW polarizations and speed, as well as detector geometry.
The specific definition and normalization chosen here are intended to facilitate generalization of the analysis beyond GR, and are not necessarily standard (see, e.g., Sect.\ 5.3 of \cite{Romano2017} for a review of these functions and their properties).

Because the noise in different instruments will generally be statistically independent \cite{gw150914detchar,Covas2018}, by cross-correlating the output of a pair of detectors, one may directly measure the {\em signal} cross-correlation of \eq{crosscor_isotropic}, and hence infer the polarization power spectra $\PSD_{\lp\lp'}(f)$ (as proposed by \cite{Grishchuk1976,Michelson1987}, and studied in multiple works since).
In a theory that allows for $N$ independent polarizations, there will be up to $N(N+1)/2$ different $S_{\lp\lp'}$ terms (only $N$ if the correlation matrix is diagonal), and at least as many detector pairs (``baselines'') will be needed to break all degeneracies between them.

\subsection{Energy density} \label{sec:energy}

Searches for a stochastic gravitational-wave background attempt to
measure the Universe's total energy density in gravitational waves as
a function of frequency. 
However, inferring this quantity from direct observables requires theoretical assumptions.
Furthermore, the equivalence principle precludes being
able to localize energy density in gravitational waves, so this is
in fact an \emph{effective} energy density.
We elaborate on these important points below, and sketch the general procedure to link the effective GW energy density to observables at the detector in (almost) any given theory.
Concrete examples of how to apply this are provided in \sect{examples}.

With an eye to cosmology, the quantity of interest in stochastic searches is usually chosen to be the log-fractional spectrum of the effective GW energy density \cite{Michelson1987, Christensen1990, Christensen1992, Flanagan1993, Allen1999}, 
\beq \label{eq:ogw}
\ogw(f) \equiv \frac{1}{\rc} \frac{\infd \rgw}{\infd \ln f}\, ,
\eeq
with $\rgw$ the effective GW energy density as a function of frequency, and $\rc$ the critical density required to close the universe,
\beq \label{eq:rhocrit}
\rc \equiv \frac{3c^2 H_0^2}{8\pi G}\, ,
\eeq
where $H_0$ is the present Hubble parameter~\cite{Allen1999}.
Presenting results of a stochastic background search in terms of this quantity facilitates their cosmological interpretation.
More importantly, using an energy density (however parametrized) allows for direct comparison with theoretical models:
in order to predict the properties of the GW background, one computes the typical GW power emitted by the system of interest (e.g., compact binaries, cosmic strings, or primordial fluctuations) and then obtains an {\em energy} spectrum by incoherently adding many such contributions (e.g., using the quadrupole formula with merger rates from population synthesis) \cite{Starobinsky1979, Easther:2006vd, Barnaby:2011qe,  Cook:2011hg, Turner:1996ck, Easther:2006gt, Kamionkowski:1993fg, Gasperini:1992em, Gasperini:1993hu, Gasperini:2016gre,Caprini2018, Zhu:2012xw, Marassi:2011si, Lasky:2013jfa, Rosado:2012bk, Zhu:2011pt, Marassi:2009ib, Buonanno:2004tp, Sandick:2006sm, Kibble:1976sj, Damour:2004kw, Siemens:2006yp, Sarangi:2002yt,Maggiore2000a}. 

However, GW detectors do not measure the effective physical GW energy density, but rather the {\em amplitude} of the waves at each instrument.
In particular, searches for a stochastic background are sensitive to the (incoherent) strain amplitude power, \eq{correlation}.
This will remain true for future detection methods, like space missions or pulsar timing.
In the case of ground-based observatories, as outlined in \sect{detection}, the stochastic strain amplitudes are probed through the cross-correlation of detector outputs across a network, \eq{crosscor}.
Thus, whatever the detection method, we will need an object that relates gravitational-wave amplitudes to
energies---a mapping that is theory-dependent.

\begin{figure*}[tb]
  \begin{center}
\ifdefined\myext
  \tikzsetnextfilename{relations}
  \tikzstyle{block} = [rectangle, draw, fill=blue!20,
  text width=8em, text centered, rounded corners,
  minimum height=4em]
\tikzstyle{arrowtext} = [above, text width=5em, text centered, minimum height=4em]
\tikzstyle{singlearrow} = [draw, arrows={-Latex[length=2mm]}]
\tikzstyle{dashdoublearrow} = [draw, dashed,
  arrows={Latex[length=2mm]-Latex[length=2mm]}]
\begin{tikzpicture}[node distance = 5cm, auto]
  \node [block] (detcrosscorr)
    {Detector cross-correlation
      $\langle \hf_{\ifo}^*(f) \hf_{\ifo'}(f') \rangle$};
  \node [block, right of=detcrosscorr] (pol2pt)
    {Polarization basis two-point function
      $\langle \hf_{A}^*(\fvec{k}) \hf_{A'}(\fvec{k}') \rangle$};
  \node [block, right of=pol2pt] (rhogw)
    {GW effective energy density
      $\rgw(\fvec{x}) = T^{\text{eff}}_{00}$};
  \node [block, right of=rhogw] (omegagw)
    {Fractional cosmological GW energy density
      $\ogw(f)$};
  \path [singlearrow] (pol2pt) --
  node[arrowtext] {\eq{crosscor}, depends on dispersion relation}
  (detcrosscorr);
  \path [singlearrow] (pol2pt) --
  node[arrowtext] {\eq{rgw-k-space-double-integ} \& \eqref{eq:tgw},
    depends on kinetic matrix} (rhogw);
  \path [singlearrow] (rhogw) --
  node[arrowtext] {\eq{ogw} \& \eqref{eq:drgw-dlnf-def},
    definition} (omegagw);
  \draw[dashed] (detcrosscorr.south) to[out=-10, in=190]
  node[arrowtext, below, text width=12em]
  {\eq{result}, when several assumptions are satisfied}
  (omegagw.south);
\end{tikzpicture}
\else
  \ifx\relsstandalone\undefined
    \tikzstyle{block} = [rectangle, draw, fill=blue!20,
  text width=8em, text centered, rounded corners,
  minimum height=4em]
\tikzstyle{arrowtext} = [above, text width=5em, text centered, minimum height=4em]
\tikzstyle{singlearrow} = [draw, arrows={-Latex[length=2mm]}]
\tikzstyle{dashdoublearrow} = [draw, dashed,
  arrows={Latex[length=2mm]-Latex[length=2mm]}]
\begin{tikzpicture}[node distance = 5cm, auto]
  \node [block] (detcrosscorr)
    {Detector cross-correlation
      $\langle \hf_{\ifo}^*(f) \hf_{\ifo'}(f') \rangle$};
  \node [block, right of=detcrosscorr] (pol2pt)
    {Polarization basis two-point function
      $\langle \hf_{A}^*(\fvec{k}) \hf_{A'}(\fvec{k}') \rangle$};
  \node [block, right of=pol2pt] (rhogw)
    {GW effective energy density
      $\rgw(\fvec{x}) = T^{\text{eff}}_{00}$};
  \node [block, right of=rhogw] (omegagw)
    {Fractional cosmological GW energy density
      $\ogw(f)$};
  \path [singlearrow] (pol2pt) --
  node[arrowtext] {\eq{crosscor}, depends on dispersion relation}
  (detcrosscorr);
  \path [singlearrow] (pol2pt) --
  node[arrowtext] {\eq{rgw-k-space-double-integ} \& \eqref{eq:tgw},
    depends on kinetic matrix} (rhogw);
  \path [singlearrow] (rhogw) --
  node[arrowtext] {\eq{ogw} \& \eqref{eq:drgw-dlnf-def},
    definition} (omegagw);
  \draw[dashed] (detcrosscorr.south) to[out=-10, in=190]
  node[arrowtext, below, text width=12em]
  {\eq{result}, when several assumptions are satisfied}
  (omegagw.south);
\end{tikzpicture}
  \else
    \includegraphics[width=\textwidth]{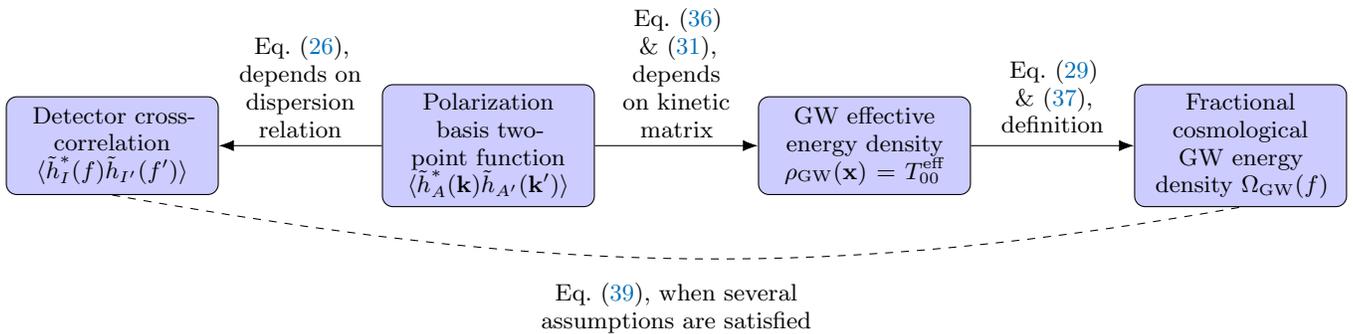}
  \fi
\fi
  \end{center}
  \caption{Key quantities appearing in stochastic searches, and how
    they are related to each other.  The relationships between them
    are theory-dependent.  The primary observable is the detector
    cross-correlation, but inferences are often stated in terms of the
    fractional cosmological GW energy density, or a parametrization
    thereof.  Arrows point from more fundamental quantities to derived
    quantities.}
  \label{fig:quantities}
\end{figure*}

The frequency-domain
\emph{effective stress-energy tensor} (ESET) for gravitational waves lets us
translate between the more accessible two-point amplitude correlation function, 
\eq{correlation}, [or the two-detector-output cross-correlation,
\eq{crosscor}] and the GW contribution to the energy density, \eq{ogw}.
In GR the ESET is given by a simple expression first derived by Isaacson \cite{Isaacson1968a, Isaacson1968b} (see \sect{gr} below), which enables stochastic searches to be parametrized directly in terms of $\ogw(f)$ \cite{Michelson1987, Christensen1990, Christensen1992, Flanagan1993, Allen1999}.
Interestingly, the same relationship has been assumed to hold in most stochastic GW data analysis schemes that allow for departures from GR \cite{Nishizawa2009, Nishizawa2010, Nishizawa2013, Chamberlin2012, Yunes2013,Cornish:2017oic,OBeirne:2018slh}, even though the Isaacson formula will not necessarily hold in arbitrary theories \cite{Stein2011}.
Using the Isaacson formula when inappropriate will lead to a
mistranslation between detector cross-correlations and GW energy densities.
This is not only misleading, but can also lead to errors when deriving
constraints on theories from observations.

In the context of any specific theory of gravity, the ESET can be derived directly from the
action.  The ESET is given by a space-time average of the variation of
the second-order perturbation of the action with respect to the
background (inverse) metric~\cite{Stein2011},
\beq \label{eq:tgw}
\tgw^\text{eff}_{ab} = \llangle[\Bigg]{-\frac{2}{\sqrt{-g^{(0)}}} \frac{\delta \act{}^{(2)}_{\rm eff}} {\delta g^{(0)ab}}}\rrangle[\Bigg]\, ,
\eeq
where the double angular brackets $\llangle \cdot \rrangle$ indicate
an averaging procedure over a spacetime region on the order of several
wavelengths (e.g.~Brill-Hartle averaging, though other 
procedures~\cite{1996GReGr..28..953Z} agree when there is a separation
of length scales).  We briefly summarize the approach here;
we refer the interested reader to \cite{Stein2011} for more exposition.

The second-order Lagrangian ${\cal L}_{\rm eff}^{(2)}$ is obtained
from the action $\act[g,\psi]$ after perturbing the metric $g_{ab}$ and
other dynamical fields $\psi$ via
\begin{align}
  \label{eq:perturb-fields}
  g_{ab} &= g_{ab}^{(0)} + \epsilon h_{ab}^{(1)}+\epsilon^{2} h_{ab}^{(2)}+\mathcal{O}(\epsilon^{3}) \,, \\
  \psi &= \psi^{(0)} +\epsilon \psi^{(1)} + \epsilon^{2} \psi^{(2)} + \mathcal{O}(\epsilon^{3}) \,,
\end{align}
and collecting terms in the action order-by-order in the small parameter $\epsilon$.  This gives the expansion
\begin{align}
  \label{eq:spert}
    \act[g,\psi] ={}& \act^{(0)}[g^{(0)},\psi^{(0)}] \\
    &{}+ \epsilon \act^{(1)}[h^{(1)},\psi^{(1)}; g^{(0)}, \psi^{(0)}]\nn\\
    &{}+ \epsilon^{2} \act^{(2)}[h^{(1,2)},\psi^{(1,2)}; g^{(0)}, \psi^{(0)}] + \mathcal{O}(\epsilon^{3})
    \, ,\nn
\end{align}
where $h^{(1,2)}$ means both $h^{(1)}$ and $h^{(2)}$ are present.
At order $\epsilon^{0}$, the action $\act^{(0)}$ generates the ordinary
nonlinear background equations of motion for $g^{(0)}$ and
$\psi^{(0)}$.  At order $\epsilon^{1}$, the action $\act^{(1)}$ is purely
a ``tadpole'' term which vanishes when $(g^{(0)},\psi^{(0)})$ are on
shell, and therefore does not contribute to any equations of motion.
The same is true for the second-order perturbations
$(h^{(2)},\psi^{(2)})$, which appear linearly in $\act^{(2)}$.
However, $(h^{(1)},\psi^{(1)})$ appear quadratically in $\act^{(2)}$: 
the quadratic action $\act^{(2)}$ then generates the linear equations of
motion for $(h^{(1)},\psi^{(1)})$ when varied with respect to
$(h^{(1)},\psi^{(1)})$;  at the same time, the variation with respect
to $g^{(0)}$ will be a quadratic functional of $(h^{(1)},\psi^{(1)})$,
and results in the ESET.

From now on we drop the order-counting superscript, letting
$h=h^{(1)}$, since we will not encounter $h^{(2)}$.
In a local Lorentz frame whose time direction is aligned with the
Hubble flow, we can define the position-space
effective GW energy density as
\begin{align}
  \rgw \equiv T^{\text{eff}}_{00}[h,h] \,,
\end{align}
where the double argument $[h,h]$ is just to remind us
that $T^{\text{eff}}$ is a quadratic functional of $h$. To use \eq{ogw}, we want
$\rgw$ in momentum space, so we need to make use of a plane-wave
expansion like Eq.~\eqref{eq:h_expansion}.  The result will always be a
momentum-space integral of the form
\begin{multline}
  \label{eq:rgw-k-space-double-integ}
  \rgw(\fvec{x}) = \int \dk \dk' \mathcal{Q}^{abcd}(\fvec{k},\fvec{k}') \times \\
  \left\langle \hf_{ab}^*(-\fvec{k}) \thinspace \hf_{cd}(\fvec{k}') \right\rangle
  e^{i(\fvec{k} + \fvec{k}')\cdot \fvec{x}} \,,
\end{multline}
where the (gauge-dependent) tensor $\mathcal{Q}^{abcd}$ encodes information about the
kinetic matrix of the theory in momentum space, and we have used \eq{ftconj} to write $\hf_{ab}(\fvec{k})=\hf^*_{ab}(-\fvec{k})$.
Notice that here we have replaced the spacetime averaging of Eq.~\eqref{eq:tgw} with
ensemble averaging, based on assumption \aref{ergodic} ergodicity.
When the two-point function $\langle\hf_{A}^{*}(\fvec{k})\hf_{A'}(\fvec{k'})\rangle$ is of the
form of \eq{correlation}, the double integral will collapse to
a single integral, and the physical energy density will be related to
the power spectral density $\PSD_{AA'}(\fvec{k})$, with some potentially
nontrivial frequency dependence arising from $\mathcal{Q}^{abcd}$ (we will
see several examples below).

When this double integral collapses to a single integral, we can then
define the fractional energy density per frequency bin via
\begin{align}
  \label{eq:drgw-dlnf-def}
  \rgw = \int \frac{\infd \rgw}{\infd f} \infd f = \int \frac{\infd \rgw}{\infd \ln f} \frac{\infd f}{f} \,.
\end{align}
With this definition of $\infd \rgw/ \infd \ln f$, and the relationship between
the energy density Eq.~\eqref{eq:rgw-k-space-double-integ} and a
two-point function like Eq.~\eqref{eq:correlation}, it will be
possible to relate the power spectral density $\PSD_{AA'}(\fvec{k})$ to
the cosmological fractional energy density $\ogw(f)$,
Eq.~\eqref{eq:ogw}.
The relationships between all these key quantities are illustrated in
Fig.~\ref{fig:quantities}.

Once we have this, we may work directly with $\ogw(f)$; in
particular, data analysis searches usually assume a power-law model like
\beq \label{eq:powerlaw}
\ogw(f) = \Omega_0 \left(\frac{f}{f_0}\right)^\alpha,
\eeq
for some spectral index $\alpha$, and $\Omega_0$ the characteristic amplitude
at some arbitrary reference frequency $f_0$. This is how LIGO generally
parametrizes its searches, e.g.\ \cite{stochastic2017}; for a discussion of the validity of this parametrization, see~\cite{Callister:2016ewt}.

\section{Example theories} \label{sec:examples}

In this section, we show how different gravitational theories imply
different functional relations between the effective fractional
energy density spectrum, $\ogw(f)$ in \eq{ogw}, the strain cross-power
spectrum, $\PSD(f)$ in \eq{correlation}, and, consequently, the
cross-correlation between detector outputs, \eq{crosscor}
As discussed in \sect{detection}, this last quantity is the relevant
observable for ground-based instruments, on which we focus.  The
relationships between all the key quantities are illustrated in
Fig.~\ref{fig:quantities}.  Along the way, we also discuss the
expected statistical properties of the polarization amplitudes under each
framework, as required purely by the basic structure of the theory (that is, not considering specific source models). 

We first demonstrate the procedure by rederiving the
standard GR expressions from the Einstein-Hilbert action
(Sec.~\ref{sec:gr}), and then offer a series of beyond-GR examples for
which the analogous result is different: 
we consider the case of Chern-Simons gravity, a
theory which is not parity-symmetric (\sect{cs}); this is followed by Brans-Dicke gravity, the prototypical example of a scalar-tensor
theory (Sec.~\ref{sec:st}); finally, we study Fierz-Pauli gravity
(Sec.~\ref{sec:mg}), in which the graviton is endowed with a mass.
The last two examples support nontensorial modes of the metric perturbation (see \sect{decomp}).

For all the examples we consider, we find it reasonable to simplify our equations by assuming the stochastic background is \aref{gaussian} Gaussian, \aref{ergodic} ergodic, \aref{stationary} stationary and \aref{isotropic} isotropic, with \aref{indepsky} no correlation between different sky locations.
In all cases, then, we find that we can write the cross-correlation between the output of two ideal differential-arm detectors $I$ and $I'$ in the form
\begin{align} \label{eq:result}
\left\langle \hf_{\ifo}^*(f) \hf_{\ifo'}(f') \right \rangle &= \frac{3 H_0^2}{4\pi^2|f|^3} \delta(f-f') \\
&\times \sum_{\lp} \Xi_{\lp}(f)\, \Omega_{\lp}(f)\, \Gamma^{\lp}_{\ifo\ifo'}(f)\, , \nonumber
\end{align}
where the sum is over some polarization basis $\lp$ that diagonalizes the kinetic matrix of the theory.
Here the $\Gamma^{\lp}_{\ifo\ifo'}(f)$'s are the generalized overlap reduction functions of \eq{overlap}, $ \Omega_{\lp}(f)$ is the effective fractional energy spectrum in polarization $\lp$ defined by analogy to \eq{ogw}, and $\Xi_{\lp}(f)$ is a model-dependent factor encoding deviations from GR.
In Einstein's theory, $\Xi_{\lp}(f)=1$ for tensor polarizations and vanishes otherwise, as we show below.

Many of the results in this section are derived on a flat background, and will therefore be erroneous in a
cosmological setting. However,
because of the vast separation of scales between the gravitational
wavelength $\lambda_{\text{GW}}$ and the Hubble parameter today
$H_{0}$, the error between the flat space results and the
cosmologically-correct results will be of fractional order
$\mathcal{O}(\lambda_{\text{GW}} H_{0} / c)$.  This correction has
been explicitly computed in GR~\cite{Bonga:2017dlx}, and while we are
not aware of the same computation in beyond-GR theories, it should
remain true as long as the theory of gravity respects the separation
of scales.

\subsection{General relativity} \label{sec:gr}

The vacuum Einstein field equations can be derived from the Einstein-Hilbert (EH) action,
\beq \label{eq:sgr}
\act_{\rm EH} = \kappa \int \infd \fvec{x} \sqrt{-g} R~,
\eeq
where $\kappa=c^4/(16\pi G)$, $g$ is the determinant of the metric $g_{ab}$, and $R$ is the Ricci scalar \cite{MTW}.
We may now expand the metric around some background, \mbox{$g_{ab} =
g^{(0)}_{ab} + \epsilon h_{ab} + \ldots\, $}, as in \eq{perturb-fields}.  The source-free linearized
equations of motion, on a flat background (so that Riemann vanishes),
and in the transverse-traceless gauge ($\nabla_a h^{ab}=0$ and $h^a{}_a=0$), take the simple form
\beq \label{eq:eomgr}
\Box h_{ab} = 0,
\eeq
where $\Box\equiv\nabla^a\nabla_a$ is the d'Alembertian with respect to the background metric.
Equation \eqref{eq:eomgr} leads to the standard geometric optics approximation to GW propagation, from which it follows that GWs show no birefringence and always propagate at the speed of light.

Focus now on the second-order perturbation of \eq{sgr}.
On a flat background, the second-order Lagrangian density is given
by~\cite{Stein2011, MacCallum1973}
\begin{align} \label{eq:lmt}
{\cal L}_{\rm GR}^{\left(2\right)} = \kappa \sqrt{-g} &\left[\frac{1}{2}
\left(\nabla_a \htr^{bc}\right) \left(\nabla_b
\htr^a_{~c}\right) - \frac{1}{4} \left(\nabla_a
\htr_{cd}\right) \left(\nabla^a \htr^{cd}\right) \right.
\nonumber \\ &\left. + \frac{1}{8} \left(\nabla_a \htr\right)
\left(\nabla^a \htr\right)\right],
\end{align}
where all derivatives are taken with respect to the background metric $g_{ab}$, and $\htr_{ab} \equiv h_{ab} - g_{ab} h^c_{~c}/2$ is the trace-reversed metric perturbation.  
This piece of the Lagrangian density corresponds to $\act^{(2)}$ in \eq{spert}.

\newcommand{\tgr}{T^{\rm (GR)}}

We now apply the transverse-traceless gauge conditions and evaluate the perturbations on-shell (that
is, we enforce the first-order equations of motion).  Then, varying with
respect to the inverse background metric, $g^{ab}$, as in \eq{tgw}, we obtain, far away from
sources,
\beq \label{eq:isaacson}
\tgr_{ab} = \frac{c^4}{32\pi G} \llangle[\big] \nabla_{a} h_{cd} \nabla_{b} h^{cd} \rrangle[\big] \,.
\eeq
This is the well-known expression for the effective stress-energy carried by a gravitational wave, first derived by Isaacson (and, consequently, known as the {\em Isaacson formula}) \cite{Isaacson1968a, Isaacson1968b}.

We may now use \eq{isaacson} to relate $S(f)$ and $\ogw(f)$, as outlined in \sect{energy}.
The Isaacson expression implies that, in a local Lorentz frame,
\beq
\rgw = \tgr_{00} = \frac{c^2}{32\pi G} \llangle[\big] \partial_t h_{ij}
\thinspace \partial_t h^{ij} \rrangle[\big]\, ,
\eeq
where we have used the fact that the transverse-traceless metric perturbation will be purely spatial.
Plugging in the plane-wave expansion of \eq{h_expansion}, using the
reality condition of \eq{ftconj}, and invoking \aref{ergodic} ergodicity, we may rewrite this as
\begin{align} \label{eq:rgw_gr_plane}
\rgw = \frac{-c^2}{128\pi^3 G} \int \dk \dk' \omega
\omega' &\left\langle \hf_{ij}^*(-\fvec{k})\, \hf^{ij}(\fvec{k}') \right\rangle \nonumber\\
&\times e^{i(\fvec{k} + \fvec{k}')\cdot \fvec{x}}.
\end{align}
This means that, in GR and in our gauge, $\rgw$ takes the form of \eq{rgw-k-space-double-integ} with
\begin{align}
  \mathcal{Q}^{abcd}_{\text{GR}} = \frac{-c^2}{128\pi^3 G} g^{ac}g^{bd} \omega \omega' \,.
\end{align}

It is convenient at this point to expand the Fourier amplitudes into polarizations.
Because GR is parity-symmetric, in this theory all modes are generated and propagate equally, so one is free to choose between linear and circular polarizations; however, working with the former is slightly simpler because the corresponding polarization tensors, Eqs.\ \eqref{eq:plus} and \eqref{eq:cross}, are real-valued.
Then, summing over $\lp$, $\lp'\in \{+,\times\}$,
\begin{align} \label{eq:rgw_gr_int}
\rgw = \frac{-c^2}{128\pi^3 G} &\int \dk \dk' \omega \omega' \left\langle \hf_{\lp}^*(-\fvec{k}) \thinspace \hf_{\lp'}(\fvec{k}') \right\rangle \nonumber \\
& \times e^\lp_{~ij} \thinspace e^{\lp'ij} ~ e^{i(\fvec{k} + \fvec{k}')\cdot \fvec{x}} .
\end{align}

We now use the fact that the Fourier amplitudes are given by a random process to simplify our expression for $\rgw$ via \eq{correlation}.
Following common practice and for the sake of simplicity, we will assume that the stationary Gaussian background is also \aref{isotropic} isotropic and \aref{equipartition} unpolarized, with equal contributions from the linear polarizations.
Letting the total PSD in tensor polarizations be $\PSD_t \equiv \PSD_+ + \PSD_\times$ with $\PSD_+= \PSD_\times = \PSD_t/2$, this means
\beq \label{eq:p_gr}
\PSDsky_{\lp\lp'}(f,\dir) = \frac{1}{8\pi}\delta_{\lp\lp'} \PSD_t(f)\, .
\eeq
Again, the assumption \aref{equipartition} of equipartition is justified because GR conserves parity.
The correlation of the Fourier polarization amplitudes, \eq{correlation_isotropic}, then becomes
\beq \label{eq:correlation_unpol}
\left\langle \hf^*_{\lp}(\fvec{k}) \hf_{\lp'}(\fvec{k}') \right\rangle \hspace{-1pt} = \hspace{-1pt}
\frac{1}{16\pi} \delta(f-f') \delta(\dir-\dir') \delta_{\lp\lp'} \PSD_t(f) .
\eeq
With this in place, and noting that \eq{lp_norm} implies $e_{\lp ij}e^{\lp ij}=4$ when summing over $A=\{+,\times\}$, the effective energy density of \eq{rgw_gr_int} simplifies to
\beq \label{eq:rgw_gr}
\rgw = \frac{\pi c^2}{4 G} \int_0^\infty \PSD_t(f) f^2~\infd f \, .
\eeq
Comparing with Eq.~\eqref{eq:drgw-dlnf-def}, we can immediately read off $\infd\rgw/\infd\ln f$, and then, from the definition of $\ogw(f)$, \eq{ogw}, we conclude
\beq \label{eq:sh_gr}
\PSD_t(f) = \frac{3 H_0^2}{2\pi^2 |f|^3} \ogw(f)\, .
\eeq

As discussed in \sect{detection}, the actual observable for stochastic-background searches in data from ground-based observatories is the cross-correlation between the outputs of pairs of detectors.
For an isotropic background, this is  given by \eq{crosscor_isotropic}, which can be written in terms of the  fractional energy density by means of \eq{sh_gr}:
\beq \label{eq:crosscor_gr}
\left\langle \hf_{\ifo}^*(f) \hf_{\ifo'}(f') \right \rangle = \frac{3
H_0^2}{8\pi^2 |f|^3} \delta(f-f') \thinspace \ogw(f) \thinspace \Gamma^{\rm t}_{\ifo\ifo'}\, ,
\eeq
where we have defined the total tensor overlap-reduction function as 
$\Gamma^{\rm t}_{\ifo\ifo'} \equiv \Gamma^{++}_{\ifo\ifo'}(f) +
\Gamma^{\times\times}_{\ifo\ifo'}(f)$.
This is the desired expression relating the observable strain cross-correlation to the fractional effective-energy density spectrum, that will be predicted by theory.
\eq{crosscor_gr} is used in most LIGO and Virgo searches for a stochastic background, via parametrizations like the $\ogw(f)$ power-law of \eq{powerlaw}.
Comparing to \eq{result}, and recalling $\ogw = \Omega_{+} +
\Omega_{\times}$ with $\Omega_{+}=\Omega_{\times}=\ogw/2$, we see that
in GR, $\Xi(f)=1$ for tensor polarizations, and vanishes otherwise, as expected.

\subsection{Chern-Simons gravity} \label{sec:cs}

Chern-Simons (CS) theory is an extension of GR with
motivations ranging from anomaly-cancellation in curved spacetime,
low-energy limits of both string theory and loop quantum gravity,
effective field theory of inflation, and more~\cite{Delbourgo:1972xb,
  Eguchi:1976db, AlvarezGaume:1983ig, Green:1984sg, Ashtekar:1988sw,
  Campbell:1990fu, Campbell:1992hc, Jackiw2003, Weinberg:2008hq,
  Taveras:2008yf, Alexander2008, Mercuri:2009zt}.  The theory is
characterized by the presence of a parity-odd, axion-like scalar
field, which couples to curvature through a parity-odd interaction
(see~\cite{Alexander2009} for a review).  The ESET in this
theory was derived in~\cite{Stein2011}, in an asymptotically-flat
spacetime and approaching future null infinity ($\mathscr{I}^{+}$).  As noted before, by
promoting flat-space results to a cosmological setting, we are making
an extremely small error of fractional order
$\mathcal{O}(\lambda_{\text{GW}}H_{0}/c)$.
Below, we provide a sketch of this derivation and show what the result implies for the stochastic background.

As a consequence of its lack of parity symmetry, CS gravity generally predicts birefringent propagation and generation of the metric perturbations, so that one of the circular tensor polarizations is amplified at the expense of the other \cite{Alexander2005}.
Consequently, as is true for any theory lacking parity symmetry, it is not appropriate to assume that the stochastic background is unpolarized \cite{Contaldi2008}.
Furthermore, as we will see, the nondynamical version of the theory predicts an expression for the effective GW stress-energy different from the Isaacson formula of \eq{isaacson}, and consequently differs from GR via a factor of $\Xi(f) \neq 1$ in \eq{result}.

\newcommand{\Scs}{\act_{\rm CS}}
\newcommand{\Sint}{\act_{\rm int}}

In the absence of matter, CS gravity is given by the Einstein-Hilbert
action of \eq{sgr}, plus terms describing the axion-curvature coupling
($\Sint$), and dynamics ($\act_\vartheta$) of the scalar field
$\vartheta$ \cite{Jackiw2003, Alexander2009},
\begin{align}
\label{eq:scs_all}
\Scs &= \act_{\rm EH} + \Sint + \act_{\rm \vartheta}\, ,\\
\label{eq:scs}
\Sint &= \frac{\alpha}{4} \int \infd \fvec x \sqrt{-g}\,\vartheta\,{\pont} \,,\\
\label{eq:scs_theta}
\act_\vartheta &= -\frac{\beta}{2} \int \infd \fvec{x} \sqrt{-g}\, g^{ab} \left(\nabla_a \vartheta \right) \left(\nabla_b \vartheta \right)
\,.
\end{align}
In the above, $\alpha$ is the constant determining the coupling of the CS field to the gravitational sector, while $\beta$ controls the kinetic energy of the scalar; $\pont$ is the Pontryagin density, which is defined in terms of the Riemann tensor, $R_{abcd}$, by
\beq
\pont = \frac{1}{2} \epsilon^{abef} R_{abcd} R^{cd} {}_{ef}\, ,
\eeq
with $\epsilon_{abcd}$ the Levi-Civita tensor.
This term is parity-odd, and gives CS gravity much of its richness.

Studying the dynamics of the theory, one may show that gravitational waves in CS gravity will present the same tensor (spin-weight 2) propagating degrees of freedom as in GR \cite{Jackiw2003, Sopuerta2009}.
On a flat background and in Lorenz gauge ($\nabla_a \htr^{ab}=0$), metric perturbations follow
the first-order equations of motion~\cite{Stein2011, Alexander2009},
\begin{align} \label{eq:eomcs}
\Box \htr_{ab} = &-\frac{1}{\kappa} \tilde{T}^{(\vartheta)}_{ab} +
\frac{\alpha}{\kappa} \left[ \nabla_c \bar{\vartheta} \nabla_d \Box
\htr_{e(a}^{~}\epsilon^{cde}{}_{b)} \right. \\ 
&+ \left. \nabla_c \nabla_d \bar{\vartheta} \epsilon^c{}_{e
f(a} \nabla^f \left(\nabla_{b)}^{~} \htr^{d e} -
\nabla^d \htr_{b)}^{~~e} \right) \right],\nonumber 
\end{align}
where we split $\vartheta$ into a smooth background piece
$\bar{\vartheta}$ and a perturbation $\tilde{\vartheta}$, and
$\tilde{T}^{(\vartheta)}_{ab}$ is the stress energy sourced
quadratically by $\tilde{\vartheta}$,
\beq
\tilde{T}^{(\vartheta)}_{ab} = \beta \left[\left(\nabla_a \tilde{\vartheta}\right)\left(
\nabla_b \tilde{\vartheta}\right) - \frac{1}{2}g_{ab}\left(\nabla^c\tilde{\vartheta}
\right)\left(\nabla_c \tilde{\vartheta}\right)\right].
\eeq
Again on a flat background, CS gravity admits an approximately
traceless gauge~\cite{Stein2011}, so that $\htr_{ab}$ can be
replaced by $h_{ab}$ in these equations, as was done for GR in
\eq{eomgr}.

In the weak-coupling limit (i.e.\ $\alpha\nabla\bar{\vartheta} \ll \kappa \lambda_{\rm GW}$, for GW perturbation wavelength $\lambda_{\rm GW} = c/f$), it can be shown that the quadratic Lagrangian density corresponding to $\act^{(2)}$ in \eq{spert} can be written as \cite{Stein2011}
\beq
{\cal L}^{(2)}_{\rm CS} = {\cal L}^{(2)}_{\rm GR} + \Delta {\cal L}^{(2)}_{\rm
CS}\, ,
\eeq
where ${\cal L}^{(2)}_{\rm GR}$ is the effective Lagrangian density derived from the Einstein-Hilbert action, \eq{lmt}, and
\begin{align} \label{eq:lcs}
\Delta {\cal L}^{(2)}_{\rm CS} \equiv \frac{\alpha}{4}\sqrt{-g}\,
\epsilon^{abcd} &\left( \nabla_e \nabla_f \bar{\vartheta}
\nabla_a \htr_b^{~f} \nabla_c \htr_d^{~e} \right. \\
&+ \left. \nabla_a \bar{\vartheta} \nabla^e \htr_b^{~f}
\nabla_d \nabla_e \htr_{fc} \right) . \nonumber 
\end{align}
From this we may derive the effective GW stress-energy, and relate the
energy to the strain cross-correlation, for both the nondynamical and
dynamical versions of the theory, as outlined in \sect{energy}.

\subsubsection{Nondynamical theory}

The nondynamical version of CS gravity is obtained from Eqs.\ \eqref{eq:scs_all}--\eqref{eq:scs_theta} by setting $\beta = 0$.
This removes the dynamics of the scalar field, fixing it to some {\em a priori} value.
Furthermore, in the {\em canonical embedding} of this theory \cite{Alexander2009}, we set the field's gradient to be purely timelike in some global frame,
\beq \label{eq:ndcs_canonical}
\nabla_\alpha \bar{\vartheta} = \mu^{-1} \delta^t_{~\alpha}\, ,
\eeq
for some constant $\mu$.  When expanding $h_{ab}$ as a power series in
$\alpha$ in the weak coupling limit,
the first-order equations of motion on a flat background, \eq{eomcs},
reduce to a simple wave equation, $\Box \htr_{ab} = 0 +
\mathcal{O}(\alpha^{2})$, as in GR.  This implies that the Lorenz
gauge is compatible with synchronous gauge (i.e., we can satisfy both $\nabla^\mu h_{\mu\nu}=0$ and $h_{0\nu}=0$ in the same frame).

In spite of its name, there is nothing special about the canonical embedding other than its simplicity \cite{Alexander2009}.
Although in the following we assume this particular form for the background scalar field, the qualitative features of our result should be similar in general, possibly with extra terms stemming from any non-zero higher-order derivatives of $\bar\vartheta$.

In the canonical embedding of nondynamical CS gravity, it can be shown that the only non-GR contribution to the on-shell ESET comes from the second term in \eq{lcs}, in regions at a great distance from the source \cite{Stein2011}.
Consequently, we can write:
\beq
T^{\rm (CS)}_{ab} = \tgr_{ab} + \Delta T^{\rm (\cs)}_{ab},
\eeq
where $\tgr_{ab}$ is the Isaacson tensor from \eq{isaacson}, and $\Delta T^{\rm (\cs)}_{ab}$ is the surviving contribution from \eq{lcs}, with components
\beq
\Delta T^{\rm (\cs)}_{\mu\nu} = \frac{\alpha}{2 \mu} \llangle[\Big]
\epsilon_i{}^{jk} \nabla_{(\mu} h^{i\sigma} \nabla_{\nu)} \nabla_{k}
h_{\sigma j} \rrangle[\Big]
\eeq
in the global frame.
The corresponding non-GR energy density, $\Delta \rho^{(\cs)} \equiv \Delta T^{\rm (\cs)}_{00}$, over a flat background is
\begin{align}
\Delta \rho^{\rm (CS)} = \frac{-i\alpha}{8\pi^2 \mu c^2} &\int \dk \dk'
\omega \omega' \epsilon^{ijk} k'_k~ e^{i(\fvec{k} + \fvec{k}')\cdot \fvec{x}}\nonumber\\&\times  \left\langle \hf^{*\,\ell}_i 
(-\fvec{k})\, \hf_{\ell j}(\fvec{k'}) \right\rangle
,
\end{align}
after expanding over plane-waves in a synchronous gauge, as done in \eq{rgw_gr_plane}, and using the reality condition of \eq{ftconj} to substitute $\hf^{~\ell}_i(\fvec{k}) \rightarrow \hf^{*\,\ell}_i(-\fvec{k})$.
In the notation of \eq{rgw-k-space-double-integ},
$\mathcal{Q}^{abcd} = \mathcal{Q}^{abcd}_{\text{GR}} + \Delta\mathcal{Q}^{abcd}$,
where in dCS and in our gauge choice,
the components of the correction are given by
\begin{align}
  \Delta\mathcal{Q}^{\alpha\beta\gamma\delta} = \frac{-i\alpha}{8\pi^2 \mu c^2} g^{\beta\gamma} \epsilon^{\alpha\delta i} k'_i \thinspace \omega \omega' \,.
\end{align}

We want to expand the perturbation into polarizations, as we did for the GR case in \eq{rgw_gr_plane}.
However, it would be inconvenient to do so in terms of the linear plus
and cross modes, since these are not actual eigenmodes of the kinetic
matrix in CS gravity, and hence their amplitudes will generally be
correlated \cite{Alexander2005}.
Instead, we will work with the right- and left-handed modes of \eq{eRL}, which {\em do} diagonalize the CS kinetic matrix.
Letting $\cp \in \{\rm R,\, L\}$, then
\begin{align} \label{eq:rcs_2}
\Delta \rho^{\rm (CS)} = \frac{-i\alpha}{8\pi^2 \mu c^2} &\int \dk \dk'~ \omega
\omega'|\omega'| \epsilon^{ijk} \hat{k}_k'\, (e^{\cp~\ell}_{~~i})^*
(e^{\cp'}_{~~\ell
j})\nonumber \\
& \times \left \langle \hf^*_{\cp}(-\fvec{k}) \hf_{\cp'}(\fvec{k}') \right
\rangle \, e^{i(\fvec{k}+\fvec{k}')\cdot \fvec{x}}.
\end{align}
Here we have used the fact that, to first order, the GW dispersion relation in canonical nondynamical CS gravity is the same as in GR, so that $k_i' = |\omega'| \hat{k}_i'$.

As a consequence of the birefringence of GWs in CS gravity, it is also no longer reasonable to assume an \aref{equipartition} unpolarized background; rather, we should expect $\PSD_{\rm R}(f) \neq \PSD_{\rm L}(f)$.
 (Although in the canonical embedding there is no amplitude birefringence in GW propagation, wave generation should still be expected to break parity symmetry.).
However, we {\em are} justified in taking the two polarizations to be uncorrelated in this basis, i.e.\ $\PSD_{\rm RL}(f) = \PSD_{\rm LR}(f)=0$, which is not true in the linear basis.

With the above considerations in mind, we may write the correlation factor in terms of the PSD in each mode as in \eq{correlation_isotropic}, so that \eq{rcs_2} becomes ($\omega'\rightarrow-\omega$):
\begin{align} \label{eq:rcs_3}
\Delta \rho^{\rm (CS)} = \frac{i\alpha \pi^2}{2\mu c^3} &\int \infd f \infd
\dir |\omega|^3 \PSD_{\cp}(f)\, \delta_{\cp\cp'}
\epsilon^{ijk} \hat{k}_k \nonumber \\
&\times (e^{\cp~\ell}_{~~i})^*(e^{\cp'}_{~~\ell j})\, .
\end{align}
With the help of Eqs.\ \eqref{eq:lc_norm} and \eqref{eq:id2}, this simplifies to our final expression for the additional energy density, after integrating over the source direction $\dir$:
\beq
\Delta \rho^{\rm (CS)} = -\frac{\alpha 8\pi^3}{\mu c^3} \int_0^\infty
\left[ \PSD_{{\rm R}}(f) - \PSD_{{\rm L}}(f)\right] f^3\, \infd f . 
\eeq
Writing the GR contribution also in terms of circular polarizations and adding it to the purely-CS part, it is then straightforward to obtain the total energy density in nondynamical CS gravity:
\beq \label{eq:rcs}
\rgw = \frac{\pi c^2}{4 G} \int_0^\infty \sum_{\cp} \lambda_{\cp}(f) \, \PSD_{\cp}(f)\, f^2  \infd f\, ,
\eeq
where the sum is over circular polarizations, and for convenience  we defined
\beq
\lambda_{\cp}(f) \equiv 1 -32 \pi^2 \rlpre_{\cp} \frac{\alpha G}{\mu c^5} f\, ,
\eeq
with $\rlpre_{\rm R/L}=\pm 1$, as in \eq{id2}.
Because the energy is diagonal in the circular modes, this may  also be written as $\rgw = \rho_{\rm R} + \rho_{\rm L}$, with each term defined as the corresponding summand (pulling the sum up front) in \eq{rcs}.

Using the definition of the fractional energy density spectrum, \eq{ogw}, this means that
the strain power in each polarization can be written as
\beq
S_{\cp}(f) = \frac{3H_0^2}{2\pi^2 |f|^3} \lambda_{\cp}^{-1}(f)\, \Omega_{\cp} (f)
\eeq
where $\Omega_{\rm R/L}(f)$ represents the energy density in each polarization, defined in full analogy to \eq{ogw} such that $\ogw = \Omega_{\rm R}(f) + \Omega_{\rm L}(f)$.
The observable cross-correlation between the output of two detectors, \eq{crosscor_isotropic}, can then be written as in \eq{result}, if we choose the circular tensor polarization basis and let
\beq
\Xi_{\cp}(f) = \lambda_{\cp}^{-1}(f) \approx 1 + 32 \pi^2 \rlpre_{\cp} \frac{\alpha G}{ \mu {c^5}}f \, ,
\eeq
with the approximation being valid in the weak-coupling limit that we have been working in ($\alpha/\mu \ll \kappa c / f$).
As expected, the usual GR expression of \eq{crosscor_gr} is recovered in the limit that the coupling of the scalar field vanishes ($\alpha\rightarrow 0$), if we further assume $\PSD_{\rm R} = \PSD_{\rm L}$.

\subsubsection{Dynamical theory}

Perhaps surprisingly, the case of {\em dynamical} CS gravity is simpler for our purposes.
This is because, in the dynamical theory, the functional form of the
effective GW stress-energy tensor (about flat spacetime and with
$\nabla\bar{\vartheta} \to 0$ far away from sources) is given by the Isaacson formula of \eq{isaacson}, as in GR \cite{Stein2011}.
This notwithstanding, dynamical CS gravity still breaks parity symmetry, featuring birefringent propagation and generation of gravitational waves.
Therefore, just as in the nondynamical theory, it would not be justified to take the stochastic background to be unpolarized.
Instead, using the circular polarization states, in which the CS kinetic matrix diagonalizes, we find that in the dynamical case 
\beq \label{eq:crosscor_circular}
\left\langle \hf_{\ifo}^*(f) \hf_{\ifo'}(f') \right \rangle = \frac{3
H_0^2}{4\pi^2 |f|^3} \delta(f-f') \Omega_{\rm \cp}(f)\, \Gamma^{\rm
\cp}_{\ifo\ifo'}(f)\, ,
\eeq
with $\cp \in \{L, R\}$, but now allowing $\Omega_{L} \neq \Omega_{R}$.
Here we also have $\Xi_{\cp}(f)=1$, as in GR.
With at least two detector pairs (e.g.\ LIGO-Livingston--Virgo, and LIGO-Hanford--Virgo), it should be possible to use this to measure the energy density in each circular mode.
\eq{crosscor_circular} may also be used to parametrize a polarized background in GR, and hence probe polarized cosmological backgrounds like those predicted in \cite{Alexander2006}.

\subsection{Scalar-tensor theories} \label{sec:st}

\newcommand{\phij}{\phi}
\newcommand{\phie}{\varphi}
\newcommand{\gstj}{g}
\newcommand{\hstj}{h}
\newcommand{\ste}[1]{\underaccent{\tilde}{#1}}
\newcommand{\gste}{\ste{g}}
\newcommand{\hste}{\ste{h}}

Scalar-tensor (ST) theories are defined by the presence of one or more scalar fields that couple to the gravitational sector nonminimally.
From a field-theoretic point of view, this family of theories are a
natural extension of GR, and, as such, has been extensively
studied~\cite{Brans1961, Lee:1974pt, Damour1992, Damour1996,
  Chiba1997, Fujii:2003pa, Sotiriou:2008rp, Sotiriou:2015lxa}.
Scalar-tensor theories are also well-motivated as effective field theories encapsulating the low-energy behavior of quantum gravity completions, like string theory \cite{Damour:1994zq,Damour:2002nv,Damour:2002mi,Taylor:1988nw}, or braneworld scenarios \cite{Randall:1999vf,Randall:1999ee}.
These theories also have important applications to cosmology \cite{Faraoni:2004pi,Clifton:2011jh}.

The literature contains several formulations of ST theories, with varying degrees of generality and complexity.
For simplicity, we will focus on the most basic case, which was introduced by Brans and Dicke in an attempt to make Einstein's theory fully compatible with Mach's principle \cite{Brans1961}.
Scalar stochastic GW backgrounds have been previously studied in the
context of this theory \cite{Maggiore2000}---we revisit some of those
results here from the ESET point of view presented in \sect{energy}.

The vacuum action for Brans-Dicke scalar-tensor gravity can be expressed as
\beq \label{eq:sstj}
\act_{\rm ST} = \kappa \int \infd {\fvec x} \sqrt{-\gstj} \left[\phij R -
\frac{\obd}{\phij} \nabla^a \phij \nabla_a \phij \right],
\eeq
for a scalar field $\phij$, some constant $\obd$, and where, as before, $\kappa = c^4/(16\pi G)$ and $R$ is the Ricci scalar.
Matter will follow geodesics of the metric associated with \eq{sstj}; this representation is known as the {\em Jordan} frame of the theory.
Notice that the scalar field has a ``scaling symmetry,'' where if we
take $\phi \to C \phi$ for some nonzero real constant $C$, this constant
can be absorbed into $\kappa$.  If the scalar field asymptotes to a
constant $\phi_{0}$ far away from all sources, we can use this scaling
symmetry to change the value of $\phi_{0}$ to whatever is most
convenient for our calculations, e.g.~we can set $\phi_{0} = 1$ (note
that~\cite{Maggiore2000} chooses a different asymptotic value).

Alternatively, it is often useful to recast the ST action in a
conformal frame in which the scalar is only minimally coupled to the metric sector.
This can be achieved by defining the conformal metric:
\beq \label{eq:gste}
\gste_{ab} \equiv \frac{\phij}{\phij_0} \gstj_{ab}\, .
\eeq
In terms of this metric and a redefined scalar field $\phie$,
Brans-Dicke theory can be recovered from the action
\beq \label{eq:sste}
\ste{\act}_{\rm ST} = \kappa \int \infd {\fvec x} \sqrt{-\gste} \left[ \ste{R}
-2 \ste{\nabla}^a \phie \ste{\nabla}_a \phie \right],
\eeq
where the under-tilded quantities are to be computed using the metric of \eq{gste}.
The new scalar field $\phie$ is related to $\phi$ from \eq{sstj} by
\begin{align}
\label{eq:st_transform}
\phij/\phij_0 &\equiv e^{-2\alpha_0 (\phie - \phie_0)}\, , \\
\alpha_0 &\equiv (3+2\obd)^{-1/2}\, ,
\end{align}
where $\phie_0$ is some constant analogous to $\phij_0$.
Because of its resemblance to the Einstein-Hilbert action of \eq{sgr}, this is known as the {\em Einstein} representation of the theory.
As we will see, \eq{sste} is more convenient for theoretical
manipulations than \eq{sstj}---although it should be kept in mind that
matter follows geodesics in~\eq{sstj}, but not in~\eq{sste}.

As usual, we will perturb the Jordan metric and field to first order by letting $\gstj_{ab} \rightarrow \gstj_{ab} + \hstj_{ab}$ and $\phij \rightarrow \phij_0 + \delta \phij$, with $\hstj_{ab} \ll \gstj_{ab}$ and $\delta \phij \ll \phij_0$, like in \eq{perturb-fields}.
For convenience, we will also define
\beq
\Phi \equiv - \delta \phij / \phij_0\, .
\eeq
Equivalently, we may perturb the Einstein-frame quantities by writing $\gste_{ab}
\rightarrow \gste_{ab} + \hste_{ab}$ and $\phie \rightarrow \phie_0 + \delta
\phie$. The two perturbations will be related by the transformation of 
\eq{st_transform}, yielding to first order:
\begin{subequations} \label{eq:st_lin-transf}
\begin{align}
\label{eq:st_h} 
 \hste_{ab} &= \hstj_{ab} - \Phi \gstj_{ab}\, ,
\\
\label{eq:st_phi}
\delta \phie &= \frac{\Phi}{2\alpha_0}\, ,
\\
\label{eq:st_glin}
\gste_{ab} &= \gstj_{ab}\, .
\end{align}
\end{subequations}

Studying linearized perturbations in the Jordan frame,
it is possible to show that there exists a gauge in which the vacuum linear equations of motion reduce to simple wave equations, $\Box \hstj_{ab} = 0$ and $\Box \Phi=0$, with the trace of the perturbation satisfying $\hstj = 2 \Phi$ \cite{Lee:1974pt}.
This implies that the metric perturbation may be locally decomposed into spin-weight 2 and spin-weight 0 contributions, in the spirit of \eq{poldecomp}, such that
\beq \label{eq:st_hpols}
\tilde{\hstj}_{ab}(\fvec{k}) = \tilde{\hstj}_+(\fvec{k})e^+_{ab}(\dir)
+ \tilde{\hstj}_\times(\fvec{k})e^\times_{ab}(\dir) +
\tilde{\Phi}(\fvec{k})e^{\rm b}_{ab}(\dir)
\eeq
with the polarization tensors as given by Eqs.\ \eqref{eq:plus},
\eqref{eq:cross} and \eqref{eq:breathing}.
It is easy to check, using the linear transformations of \eq{st_lin-transf}, that in the Einstein frame this is equivalent to a gauge in which the trace-reversed Einstein metric is given by the same expression as the Jordan metric, i.e.\
\beq \label{eq:st_htr}
\ste{\htr}_{ab} \equiv \ste{h}_{ab} - \ste{h} \thinspace \eta_{ab}/2 = \hstj_{ab}\, .
\eeq
Consequently, $\ste{\htr}_{ab}$ is divergenceless ($\nabla^a \ste{\htr}_{ab} = 0$), although it is {\em not} traceless ($\ste{\htr}=\hstj = 2\Phi$).

We will now derive an expression for the GW effective stress-energy in the {\em Einstein} frame, and will then re-express this in terms of the Jordan quantities that are measurable at the detector.
The reason for this choice is that, by definition of the Einstein frame, the metric and scalar field separate in the action of \eq{sste}.
This nice feature not only makes {\em our} computations easier, but also those in the modeling of observational scenarios for the stochastic background---which will generally also offer a prediction of the energy spectrum in the Einstein frame.
In any, case there is no difference between the Jordan and Einstein energies to linear order, as given by \eq{st_lin-transf}.

From the variation of the second-order perturbation of the Einstein-frame Lagrangian density, \eq{sste}, with respect to the inverse background metric, $\gste^{\mu\nu}$, we can show, as in \sect{energy}, that the effective GW stress-energy tensor will be given by two terms:
\beq \label{eq:tst}
\ste{T}_{ab}^{\rm (ST)} = \ste{T}^{\rm (EH)}_{ab} + \Delta \ste{T}^{\rm (ST)}_{ab}.
\eeq
The first, $\ste{T}^{\rm (EH)}_{ab}$, is the contribution from the Einstein-Hilbert part of the action in \eq{sste}---this is analogous to $\tgr_{ab}$ in \eq{isaacson}, but is {\em not identical} to it due to the presence of the scalar and the necessarily different gauge choice with $\ste{\htr}=2\Phi$.
In fact, starting from the quadratic Lagrangian density of \eq{lmt},
after evaluating on shell,
it may be shown that
\beq \label{eq:tst_eh}
\ste{T}^{\rm (EH)}_{ab} = \frac{\kappa}{2} \llangle[\Big] \nabla_a
\hstj_{cd} \nabla_b \hstj^{cd} - 2 \nabla_a \Phi \nabla_b \Phi
\rrangle[\Big] \, ,
\eeq
in a synchronous gauge for $\hstj_{ab}$ (Appendix \ref{app:st}).
Recall that the metric perturbation appearing in this equation can be equivalently taken to be the trace-reversed perturbation in the Einstein frame, or the regular perturbation in the Jordan frame ($\hstj_{ab}=\ste{\htr}_{ab}$), because we are working to linear order [\eq{st_htr}].

The second contribution to the stress energy comes from the variation of
the second term in \eq{sste}, and can be shown to be
\beq \label{eq:tst_phi}
\Delta \ste{T}^{\rm (ST)}_{ab} = (3 + 2\obd)\kappa 
\llangle[\big] \nabla_a \Phi \nabla_b \Phi \rrangle[\big]\, ,
\eeq
after applying the equations of motion (Appendix \ref{app:st_eset}).
In both Eqs.\ \eqref{eq:tst_eh} and \eqref{eq:tst_phi}, we have simplified the notation by letting $\ste{\nabla} \rightarrow \nabla$, because these derivatives are taken with respect to the background metric, $\gste_{ab} = \gstj_{ab}$ to linear order [\eq{st_glin}].
Adding together the two contributions, we obtain the total Einstein
frame stress energy:
\beq \label{eq:tst_all}
\ste{T}_{ab}^{\rm (ST)} = \frac{\kappa}{2} \llangle[\big] \nabla_a \hstj_{cd}
\nabla_b \hstj^{cd} \rrangle[\big] + 2 \kappa \left(1 + \obd\right)
\llangle[\big] \nabla_a \Phi \nabla_b \Phi \rrangle[\big]\, .
\eeq
This agrees with the expression originally found in \cite{Lee:1974pt} by a different procedure.

As in previous sections, we may now expand the corresponding effective energy density, $\rgw \equiv T_{00}$, into plane-waves to obtain an expression like \eq{rgw-k-space-double-integ} with ${\cal Q}^{abcd}= {\cal Q}^{abcd}_{\rm GR} + \Delta {\cal Q}^{abcd}$ and 
\beq \label{eq:Qst}
\Delta {\cal Q}^{abcd} = -\frac{c^2 \omega\omega'}{128\pi^3 G} \left(\obd + 1\right) g^{ab}\thinspace g^{cd}\, ,
\eeq
where we have used the fact that $\Phi = g^{ab} h_{ab} / 2$, as implied by \eq{st_hpols}.
The energy density can also be written explicitly in terms of the polarization amplitudes as
\begin{align}
\rgw = - \frac{\kappa}{2c^2} \int & \dk \dk ' \omega \omega'
 e^{i(\fvec{k} + \fvec{k'})\cdot \fvec{x}} \  \times \\
& \sum_{\lp} \lambda_{\lp} \left\langle \tilde{\hstj}^*_{\lp}(-\fvec{k}) \tilde{\hstj}_{\lp}(\fvec{k})
\right\rangle \, , \nn
\end{align}
with a sum over the polarizations $\lp \in \{\rm +,\, \times,\, b\}$.
To make the notation more compact, we have also defined the auxiliary factor
\beq \label{eq:stfactor}
\lambda_\lp = \begin{cases} 
(3+2\obd) & {\rm if~} \lp = {\rm b}, \\
1 & {\rm if~} \lp = +, \times.
\end{cases}
\eeq
For more details, see Appendix \ref{app:st_rho}. 

We must now make some assumptions about the statistical properties of the Fourier amplitudes.
As before, we will assume the simplest case of \aref{isotropic} an isotropic background, with \aref{indeppol} uncorrelated polarizations and \aref{indepsky} sky-bins.
We can then use the corresponding expression for the correlations, \eq{correlation_isotropic}, to get:
\beq \label{eq:strgw}
\rgw = \frac{\pi c^2}{4 G} \sum_{\lp} \int_0^\infty \infd f\,f^2 \lambda_{\lp} S_{\lp}(f)\, .
\eeq
From the definition of the fractional energy spectrum, \eq{ogw}, this in turn implies [cf.\ \eq{drgw-dlnf-def}]
\beq
S_{\lp}(f) = \frac{3H_0^2}{2\pi^2 |f|^3} \lambda_{\lp}^{-1}  \Omega_{\lp} (f)\, ,
\eeq
where  $\Omega_{\lp}$ represents the energy content in polarization $A$, with $\ogw = \sum \Omega_{\lp}$ for $A\in\{+,\, \times,\, {\rm b}\}$, because we took the different modes to be uncorrelated.
This is justified because the kinetic matrix of the theory is diagonal for $A\in\{+,\, \times,\, {\rm b}\}$.

We may use this expression for the power spectral density in each polarization to write the observable cross-correlation between the output of a pair of  detectors ($\ifo$ and $\ifo'$).
Using the cross-correlation expression of \eq{crosscor_isotropic}, we find again that we can write this as in \eq{result}
with a summation over polarizations  $A\in\{+,\, \times,\, {\rm b}\}$, and the factor 
\beq
\Xi_{\lp}(f) = \lambda_{\lp}^{-1} = 
\begin{cases} 
(3+2\obd)^{-1} & {\rm if~} \lp = {\rm b}, \\
1 & {\rm if~} \lp = +, \times.
\end{cases}
\eeq
The GR result of \eq{crosscor_gr} is recovered, as expected, in the limit that $\obd$ becomes infinitely large.

\subsection{Massive gravity} \label{sec:mg}

\newcommand{\newMG}{f_{m}}
\newcommand{\mgratio}{\alpha}
\newcommand{\mgnullity}{\beta}
\newcommand{\lcompt}{\lambda_m}
\newcommand*{\syn}[1]{\ste{#1}}

From a field theory perspective, general relativity is nothing but the theory of a nontrivial massless spin-2 particle---the graviton.
Therefore, theories of massive gravity, which endow the graviton with a mass, are a natural (and, in some sense, the simplest) extension of Einstein's theory (see \cite{DeRham2014,Hinterbichler2011} for reviews).
In its most basic form, linearized massive gravity is given by the Fierz-Pauli (FP) action \cite{Fierz1939},
\beq \label{eq:spf}
\act_{\rm FP} = \act^{(2)}_{\rm EH} + \act_{m}\, ,
\eeq
where $\act^{(2)}_{\rm EH}$ is the quadratic piece of the Einstein-Hilbert action of \eq{sgr}, and $\act_{m}$ is the Fierz-Pauli mass term,
\beq \label{eq:sm}
\act_{m} = \frac{1}{2} \kappa \int \infd \fvec{x} \sqrt{-g}\, \mG^2 h_{ab}h_{cd}\thinspace g^{a[b}g^{c]d} \, ,
\eeq
for a graviton mass $m=\hbar \mG/c$, and where $h_{ab}$ is a linear perturbation over the background metric $g_{ab}$, as before.
For background diffeomorphism invariance, we explicitly include the
$\sqrt{-g}$ term in \eq{sm}, though the background metric in this
action should be thought of as Minkowski (yet potentially in
curvilinear coordinates).

Extending this linear theory to a more general, nonlinear one is far from trivial (for
reviews see e.g.~\cite{DeRham2014,Hinterbichler2011}).
Therefore, we will focus only on the linear theory of~\eq{spf}, and will
only comment on the relevance of the linearized analysis for the nonlinear completion at the end of
this section. 
Until then, we will write ``massive gravity'' to
mean Fierz-Pauli theory.

Massive gravity has many interesting features, including the fact that it supports five independent GW polarizations corresponding to the helicity states available to a massive particle: two tensor modes (helicity $\pm2$), two vector modes (helicity $\pm1$), and one scalar mode (helicity $0$).
Over a flat background, these degrees of freedom propagate following the Klein-Gordon equations of motion describing a massive graviton,
\beq \label{eq:mg_box}
\left(\Box - \mG^2\right) h_{ab} = 0\, ,
\eeq
and are divergenceless and traceless,
\begin{subequations} \label{eq:mg_constraints}
\begin{align} \label{eq:mg_div}
\nabla_a h^{ab}=0\, ,
\\ \label{eq:mg_trace}
h = g^{ab} h_{ab} = 0\, .
\end{align}
\end{subequations}
These three equations follow from the variation of \eq{spf} with respect to the inverse metric perturbation $h^{ab}$ \cite{DeRham2014,Hinterbichler2011}, and contain all relevant properties of GWs in this theory.
Equation~\eqref{eq:mg_box} immediately gives the dispersion relation
$\omega^2=c^2 (\normk^2 + \mG^2)$.

Before proceeding, we must discuss the length scales which appear in
this calculation.  Around a flat background, there are only two length
scales of importance: the wavelength of radiation, $\lambda_{\rm GW}$,
and the graviton's Compton wavelength, $\lcompt = h/mc$.
Generally speaking, the relevance of corrections to GR due to a
nonvanishing graviton mass will depend on the value of the ratio of
these two,
\beq \label{eq:mg_ratio}
\mgratio (f) \equiv \frac{\lambda_{\rm GW}}{\lcompt} = \frac{c \mG}{2\pi f}\, ,
\eeq
or, equivalently, the ratio of the norm of the wave's spatial wavevector to its angular frequency,
\beq \label{eq:mg_ratio_2}
\mgnullity (f) \equiv  \frac{\normk c}{2\pi f}
= \sqrt{1-\mgratio^2}\, .
\eeq
This last quantity is just the graviton group velocity in natural units, which is the same as the ratio of the speed of light to the graviton phase velocity ($\beta = v_g / c = c / v_p$).
We should expect to recover GR results for vanishing graviton mass, when $\mG\rightarrow0$ and, consequently, $\mgratio \rightarrow 0$ and $\mgnullity \rightarrow 1$.
Note that for propagating GW modes, we must have $\mgratio < 1$, and
consequently, $\mgnullity$ is real-valued.

When we move to the cosmological setting (or a more general curved
background), there is a third length scale at each point: the
curvature radius of the background, $L_{\rm BG}$.  In order for the
Brill-Hartle averaging procedure to be valid, we need a separation of
scales, $\lambda_{\rm GW}\ll L_{\rm BG}$, since the B-H average makes
errors of order $\lambda_{\rm GW} / L_{\rm BG}$ (this is clearly
satisfied when comparing the LIGO/Virgo frequency band with the cosmological
curvature radius $cH_{0}^{-1}$).  Now, in the following, we want to
keep the dependence on $\mu$, so we keep terms at the length
scale $\lcompt$.  This is only compatible with the B-H averaging
procedure if we demand the additional separation of scales
$\lcompt \ll L_{\rm BG}$.

We now return to a flat background to develop the results which we
later promote to a cosmological background.
In a generic frame (that is, without special boosts) with rectangular
coordinates, and with the $z$-axis aligned along the wave's direction of propagation, the equations of motion can be shown to restrict the components of a massive GW to be of the form (Appendix \ref{app:mg_unitary}):
\beq \label{eq:h_mg_unitary}
(h_{\mu\nu}) = \begin{pmatrix}
\mgnullity^2 h_{\rm l}  & -\mgnullity h_{\rm x} & -\mgnullity h_{\rm y}    & -\mgnullity h_{\rm l}  \\
- \mgnullity h_{\rm x} & -\frac{1}{2}\mgratio^2 h_{\rm l} + h_+ & h_\times    & h_{\rm x}  \\
- \mgnullity h_{\rm y} & h_\times    & -\frac{1}{2}\mgratio^2 h_{\rm l} - h_+ & h_{\rm y}  \\
- \mgnullity h_{\rm l} & h_{\rm x}    & h_{\rm y}    & h_{\rm l}
\end{pmatrix} ,
\eeq
for the five linear polarization amplitudes $h_{\lp}$, with $\lp \in \{\rm +, \times, x, y, l\}$.
Here, we have parametrized the single scalar mode allowed by the theory in terms of the longitudinal amplitude (rather than the breathing amplitude, or some linear combination thereof), treating it as the fundamental degree of freedom.%
\footnote{Importantly, note that our definition of the longitudinal mode follows the standard in the GW literature, and does not necessarily agree with the conventions from the massive-gravity theory literature, e.g.\ Ref.\ \cite{Hinterbichler2011} defines the longitudinal tensor as proportional to our $e^{\rm l}_{ab} - e^{\rm b}_{ab}/2$, instead of just $e^{\rm l}_{ab}$.}
It is straightforward to check that the metric of \eq{h_mg_unitary} is traceless and divergenceless, as required by the equations of motion (Appendix \ref{app:mg_unitary}).
The GR case is recovered in the limit of vanishing graviton mass, if we also re-enforce the requirements of transversality and tracelessness by setting $h_{\rm x}=h_{\rm y}=h_{\rm l}=0$.

We must now determine the functional form of the GW
effective stress-energy tensor in FP theory.
Varying the effective Lagrangian density from \eq{spf} with respect to $g^{ab}$, as in \eq{tgw}, we may write the effective GW stress-energy tensor in massive gravity as (Appendix \ref{app:mg_energy})
\beq \label{eq:tfp_all}
\tgw_{ab}^{\rm (FP)} = T^{\rm (EH)}_{ab} + \Delta T^{\rm(FP)}_{ab}.
\eeq
As in previous examples, $T^{\rm (EH)}_{ab}$
is derived from the Einstein-Hilbert piece of the action, but now
evaluated with the new on-shell condition \eq{mg_box}, rather than the
GR requirement of \eq{eomgr}.  This gives
\begin{align} \label{eq:tfp_eh}
T^{\rm (EH)}_{ab} ={}& \frac{\kappa}{2} \llangle[\Big] \nabla_a h^{cd} \nabla_b h_{cd} \rrangle[\Big] \\
&{}+ \kappa \mG^2 \llangle[\Big] h_{da} h_b{}^d
+ \tfrac{1}{4} g_{ab} h^{cd} h_{cd}  \rrangle[\Big]\, . \nonumber
\end{align}
 The second term in \eq{tfp_eh} is derived from the Fierz-Pauli mass term of \eq{sm} and reduces to
\begin{align} \label{eq:tfp}
\Delta T_{ab}^{\rm (FP)}
&= -\kappa \mG^2  \llangle[\Big] h_{da} h_b{}^d
 +\tfrac{1}{4}g_{ab} h^{cd} h_{cd} \rrangle[\Big] \,,
\end{align}
for on-shell perturbations (Appendix \ref{app:mg}).
This result includes no derivatives of the metric, as is to be expected from \eq{sm}.
Perhaps surprisingly, the mass
terms appearing \eq{tfp_eh} and \eq{tfp} exactly cancel, resulting in the same
functional form for the ESET as in GR,
\begin{align} \label{eq:tfp_sum}
  T^{\rm (FP)}_{ab} ={}& \frac{\kappa}{2} \llangle[\Big] \nabla_a h^{cd} \nabla_b h_{cd} \rrangle[\Big]
  \,.
\end{align}
This result is in agreement with one derived in~\cite{Finn:2001qi}
based on Noether's theorem on a Minkowski background (though note that
Ref.~\cite{Finn:2001qi} had a slightly different mass term, but this
difference cancels out after evaluating the ESET on shell).
Despite the fact that the two functionals have the same on-shell
expressions, the solutions $h_{ab}$ on which they will be
evaluated differ, because they satisfy different linearized equations
of motion, \eq{eomgr} vs.~\eq{mg_box}.
We caution that, as discussed at the end of this section, the Isaacson
expression \eq{tfp_sum} should not be expected to hold in a nonlinear
completion of the theory over arbitrary backgrounds.

Decomposing the metric components into plane-waves, the above expressions imply that the energy density $\rgw\equiv T_{00}$, in some frame, may be written as in \eq{rgw-k-space-double-integ} with ${\cal Q}^{abcd}= {\cal Q}^{abcd}_{\rm GR}$.
Breaking up the Fourier amplitudes into polarizations and applying all the usual assumptions \aref{gaussian}--\aref{isotropic} about the background, it may then be shown that we can use \eq{correlation_isotropic} to write the energy density in terms of the polarization PSDs as (Appendix \ref{app:mg_energy}): 
\beq \label{eq:rgw_mg}
\rgw = \frac{\pi c^2}{4G} \int_0^\infty \sum_\lp \lambda_\lp(f) S_\lp(f)\, f^2\infd f\, ,
\eeq
where the sum is over the five linear polarizations $A \in \{+, \times, {\rm x}, {\rm y}, {\rm l}\}$ as they appear in \eq{h_mg_unitary}, and we have defined
\beq \label{eq:lambda_mg}
\lambda_\lp(f) \equiv \begin{cases} 
1 & {\rm if~} \lp = +,\times , \\
\mgratio^2 & {\rm if~} \lp ={\rm x},{\rm y} ,\\
\tfrac{3}{4} \mgratio^4 & {\rm if~} \lp = {\rm l}\, .
\end{cases}
\eeq
Clearly, higher powers of $\mgratio$ will be strongly suppressed in the limit of small mass we are working in, but leave them in for now nonetheless.  
Note again that we have assumed that the polarization amplitudes of \eq{h_mg_unitary} are statistically independent because they are the fundamental degrees of freedom that diagonalize the kinetic matrix of the theory.

With this expression for $\rgw$ in hand, the definition of the fractional energy density spectrum, \eq{ogw}, then implies that
\begin{align} \label{eq:ogw_mg}
\ogw (f)
&= \frac{2 \pi^2 |f|^3}{3 H_0^2} \sum_\lp \lambda_\lp(f) S_\lp(f)\, ,
\end{align}
and, as we have done in previous sections, we may call each summand in this equation $\Omega_{\lp}(f)$,
with $\ogw(f) = \sum_\lp \Omega_\lp(f)$, so that we can write the corresponding polarization spectral density as
\beq \label{eq:psd_mg_unit}
S_\lp (f) = \frac{3 H_0^2}{2\pi^2 |f|^3} \lambda_\lp^{-1}(f)\, \Omega_\lp(f)\, .
\eeq

We now want to relate the GW energy density to the cross-correlation of the outputs of two interferometric detectors.
Instead of \eq{h_mg_unitary}, we would like to be able to write the GW as a purely spatial metric perturbation ($h_{0\nu} = 0$) in arbitrary frames (i.e.\ without the need for special boosts).
This is so we can have the perturbation be purely spatial in the proper frame of the detector, which would then  allow us to use our usual expression for the detector tensor, \eq{detector}, when computing the output of a measurement.

In GR, the required gauge freedom is afforded by diffeomorphism
invariance, which is not directly available to us in massive
gravity~\cite{DeRham2014, Hinterbichler2011}.
However, we may circumvent this restriction by introducing auxiliary
fields into the action, designed to reintroduce gauge freedom to the
theory (the so-called {\em \Stuck{} trick}).
We would then obtain a generalized version of massive gravity that
is invariant under infinitesimal coordinate transformations, and
which reduces to the usual theory after fixing to a particular
gauge (see Appendix \ref{app:mg_sync}).

We refer to the gauge that returns the FP action of \eq{spf} as the {\em unitary} gauge, as opposed to the {\em synchronous} gauge, in which the metric perturbation can take a purely-spatial form without special boosts.
In this gauge, a measurement via an interferometric detector in the small-antenna limit can be represented by the double contraction of the metric with the detector tensor of \eq{detector}, and the metric perturbation can be decomposed as in \eq{polarizations}, as explained in \sect{detection}. 

Unfortunately, the synchronous polarizations will {\em not} be statistically independent in the linear basis of \eq{polarizations}, which is commonly used in data analysis (e.g.\ \cite{Callister2017}).
In fact, the 6 polarization amplitudes in the synchronous gauge,
$\syn{h}_{A}$, can be obtained from the 5 in the unitary gauge of
\eq{h_mg_unitary}, $h_B$, via a (polarization-basis-dependent)
transformation matrix, $M_{\syn{A}}{}^{B}$, given by
\begin{align} \label{eq:h_mg_transf}
(M_{\syn{\lp}}{}^{B}) \equiv
  \begin{pmatrix}
    1 & 0 & 0 & 0 & 0 \\
    0 & 1 & 0 & 0 & 0 \\
    0 & 0 & \alpha^{2} & 0 & 0 \\
    0 & 0 & 0 & \alpha^{2} & 0 \\
    0 & 0 & 0 & 0 & -\tfrac{1}{2}\alpha^{2} \\
    0 & 0 & 0 & 0 & \alpha^{4}
  \end{pmatrix} ,
\end{align}
so that $\syn{h}_A =M_{\syn{A}}{}^{B} h_B$, and where $\syn{A} \in \{\rm +,\, \times,\, x,\, y,\, b,\, l\}$ indexes {\em synchronous} polarization amplitudes $\syn{h}_A \in \{ \syn{h}_+,\, \syn{h}_\times,\, \syn{h}_{\rm x},\, \syn{h}_{\rm y},\, \syn{h}_{\rm b},\, \syn{h}_{\rm l}\}$, while $B \in \{\rm +,\, \times,\, x,\, y,\, l\}$ indexes {\em unitary} polarization amplitudes $h_B \in \{ {h}_+,\, {h}_\times,\, {h}_{\rm x},\, {h}_{\rm y},\, {h}_{\rm l}\}$.
Had we started with a basis for the unitary metric components different than \eq{h_mg_unitary}, all our results would still apply after redefining $M_{\syn{A}}{}^{B}$ accordingly.
We provide an explicit expression for $\syn{h}_{ab}$ in terms of the unitary amplitudes in \eq{h_mg_synchronous_app} in Appendix \ref{app:mg_sync}.
The fact that the 5 unitary amplitudes determine 6 synchronous
amplitudes makes it immediately clear that the latter are not
statistically independent.

Taking advantage of the synchronous gauge to compute detector responses and taking the unitary polarizations to be uncorrelated, the cross-correlation of two detector outputs may be written directly in terms of the fractional energy spectrum for each unitary polarization via \eq{psd_mg_unit},
\begin{align} \label{eq:mg_crosscor}
\left\langle \hf_{\ifo}^*(f) \hf_{\ifo'}(f') \right \rangle &= 
\frac{3 H_0^2}{4\pi^2 |f|^3} \delta(f-f') \sum_{B} \Omega_B(f) \\
&\times \lambda^{-1}_B(f) M_{\syn{\lp}B}(f)\, M_{\syn{\lp}'B}(f)\, \Gamma^{\syn{\lp}\syn{\lp}'}_{~~\ifo\ifo'}(f)\, , \nonumber 
\end{align}
with $\lambda_B(f)$ as in \eq{lambda_mg}, $M_{\syn{A}B}$ as in \eq{h_mg_transf}, and $\Gamma^{\syn{\lp}\syn{\lp}'}_{~~\ifo\ifo'}(f)$ the generalized overlap reduction functions for the synchronous polarizations.
These functions are defined as in \eq{overlap}, with a delay factor corresponding to $v_p = c/ \beta$ [cf.~\eq{mg_ratio_2}], i.e.
\beq
\xi^{AA'}_{II'} = \frac{\Delta x_{\ifo\ifo'}}{c}\sqrt{1-\alpha^2} 
\approx \frac{\Delta x_{\ifo\ifo'}}{c} \left( 1 - \frac{1}{2} \alpha^2 \right) ,
\eeq
after expanding for small $\alpha$.
The resulting overlap-reduction functions will not be the same (even ignoring differences in normalization) as those used in existing stochastic searches beyond GR \cite{Callister2017, nongrsgwo1}, because those assume $v_p=c$.
However, we should expect that to be a good approximation as long as the extra delay in the time of flight due to the nonvanishing mass, $\delta \xi^{AA'}_{II'} \equiv -\frac{1}{2}\alpha^2 \Delta x_{\ifo\ifo'}/c$, is small with respect to the timing accuracy of the instruments.
For a treatment of overlap-reduction functions without ignoring this correction, see \cite{Nishizawa2013}. 

Regardless of whether we neglect dispersive corrections to the overlap-reduction functions or not, it turns out that, for differential-arm detectors, we have that
\beq \label{eq:mg_overlap}
\Gamma^{\syn{\lp}\syn{\lp}'}_{~~\ifo\ifo'} =
\begin{cases} 
\left(2 \delta_{\syn{\lp}\syn{\lp}'} - 1\right) \Gamma^{\rm l}{}_{\ifo\ifo'}  & \text{if $\syn{\lp}$ or $\syn{\lp}'$ in $\{\rm b, l\}$},\\
\delta_{\syn{\lp}\syn{\lp}'} \Gamma^{\syn{\lp}}{}_{\ifo\ifo'}  & \text{otherwise},
\end{cases}
\eeq
as long as differences in the phase velocities of different polarizations are negligible (which is exactly the case for the Fierz-Pauli theory).
This relation may be used to put our result of \eq{mg_crosscor} in the form of \eq{result} with
\begin{align} \label{eq:xi_mg}
\Xi_\lp (f)
&\approx \begin{cases}
1 & \text{if $\lp$ in $\{+,\times\}$},\\
\mgratio^2 & \text{if $\lp$ in $\{\rm x, y\}$},\\
\frac{1}{3}\left(2\mgratio^2 + 1\right)^2 & \text{if $\lp = {\rm l}$}. 
\end{cases} 
\end{align}
plus terms of order $\mgratio^6$ and higher.
Here, $\Xi_\lp(f)$ goes smoothly to the GR limit as $\mgratio \to 0$
(vanishing graviton mass) for the tensor and vector modes.  However,
notice that $\Xi_{l}(f) \to \frac{1}{3}$ (rather than vanishing) in
this same limit.  This is reminiscent of the vDVZ (van Dam, Veltman,
Zakharov) discontinuity~\cite{vanDam:1970vg, Zakharov:1970cc} (see
also~\cite{VanNieuwenhuizen:1973qf} for a similar effect,
and~\cite{Hinterbichler2011} for more discussion).
For interesting details on the derivation of Eqs.~\eqref{eq:mg_crosscor}--\eqref{eq:xi_mg}, we refer the reader to Appendix \ref{app:mg_crosscor}.

\subsubsection*{Relation to nonlinear massive gravity}

There is no problem in thinking of the action \eq{spf} as describing a
linear spin-2 field $h_{ab}$ on a curved background $g_{ab}$.
However, if we want $h_{ab}$ to represent metric fluctuations of the
gravitational field, the theory must have a nonlinear completion,
which is known to have several problems (see
e.g.~\cite{Hinterbichler2011, DeRham2014} for more discussion).  One
which we have already mentioned [below \eq{xi_mg}] is the vDVZ
discontinuity, by which the limit of vanishing graviton mass $\mG \to 0$
does not recover GR (e.g.,~the scalar degree of freedom does not
decouple).

Another major problem is the Boulware-Deser
ghost~\cite{Boulware:1973my}, which must be excised order-by-order in
the graviton self-interaction.  Controlling this ghost degree of
freedom to all orders is possible with a specific set of
self-interactions, known as de~Rham-Gabadadze-Tolley (dRGT) massive
gravity~\cite{deRham:2010kj}.  This has been extended to a theory of
two interacting metrics by Hassan and Rosen~\cite{Hassan:2011zd},
which has dRGT as a careful scaling limit.  Bigravity propagates one
massive and one massless spin-2 field (7 total degrees of freedom),
whereas taking the dRGT limit eliminates the massless mode (leaving
only 5 dynamical degrees of freedom, as in the linearized theory).

Indeed when expanded about a Minkowski background to linear order
(quadratic in the Lagrangian), dRGT agrees with Fierz-Pauli theory.
This might lead one to believe that the preceding FP analysis can be
directly lifted to dRGT, or even to bigravity, but this conclusion is
unwarranted.  The quadratic Lagrangian about nontrivial background-field configurations~\cite{Alberte:2011ah, Hassan:2012wr,
  Bernard:2014bfa, Bernard:2015mkk, Bernard:2015uic} can look rather
different from the simple FP Lagrangian.\footnote{%
For a special subclass of ``proportional'' background configurations
in bigravity~\cite{Hassan:2012wr, Bernard:2015uic}, two linear
combinations of the two metrics' perturbations can be combined into
the massless and massive eigenstates which diagonalize the kinetic
matrix of the quadratic Lagrangian.  In this case, the massive mode
does have a FP Lagrangian.  However, this is likely a special
case---as far as we have been able to discern, the transformation to
the mass eigenstates has not been performed for a more general
background.
}

In fact, the difference from the FP Lagrangian is crucial for the
health of such theories, because otherwise the nonlinear theories
would also exhibit problematic phenomenology, like the vDVZ discontinuity.
However, healthy nonlinear massive gravity theories are protected from
vDVZ phenomenology by the Vainshtein screening
mechanism~\cite{Vainshtein:1972sx}.  The Vainshtein mechanism leads to
a nontrivial, nonlinear field configuration (like a condensate) with a
new length scale, the Vainshtein radius.  Within this radius, the
\emph{effective} couplings for the massive degree of freedom can be
very different from what is seen when expanded about the Minkowski
background, thus reverting to the phenomenology of general relativity.

In short, the ESET for nonlinear massive gravity on a general
background (e.g.~one exhibiting Vainshtein screening) should be
considered an open problem.  It seems unlikely that the FP result
lifts to the general massive gravity result.

\section{Conclusion} \label{sec:conclusion}

The detection of a stochastic gravitational-wave background will provide a unique opportunity to study the properties of gravitational waves as they propagate through cosmological distances, and will thus be an invaluable tool to study extensions of general relativity.
Properly interpreting the theoretical implications of such a detection will require a detailed understanding of the assumptions that go into the usual searches for a stochastic background, and how the measurement process might be modified in theories beyond general relativity. 
Towards that goal, in this paper we have laid out the formalism underlying searches for stochastic signals in a generic fashion that makes it easily applicable to a large family of theories.
We have also surveyed the standard set of assumptions that go into these searches, evaluating their generic applicability, or lack thereof.

First and foremost, we find that most existing treatments of stochastic backgrounds beyond GR fail to consider possible modifications to the effective stress-energy carried by a gravitational wave of a given amplitude and frequency \cite{Nishizawa2009, Nishizawa2010, Nishizawa2013, Chamberlin2012, Yunes2013,Cornish:2017oic,OBeirne:2018slh}.
This is important because the {\em goal} of searches for stochastic backgrounds, within GR or beyond, is precisely to measure the amount of energy that exists in the form of stochastic gravitational waves.
Accordingly, data analysis strategies tend to be parametrized directly in terms of of an effective energy spectrum, \eq{ogw}.
However, this is only possible if one knows the relation between the energy density and the observables at the detector (e.g.\ the cross-correlation of strain detector outputs)---this will depend on the specific structure of the underlying theory of gravity, and in general need not be the same as in GR.
Therefore, parametrizing model-independent searches for backgrounds beyond GR as traditionally done will result in the use of a quantity that should {\em not} generally be interpreted as the GW energy density, and may thus lead to incorrect comparisons between theory and experiment.
Instead, we find it advisable to parametrize theory-agnostic searches using the power spectrum of polarization amplitudes, \eq{correlation}, which have a (mostly) model-independent interpretation.  One can always translate amplitudes into effective energies for any specific theory, as sketched in \sect{energy}.

We also reviewed the standard set of simplifying assumptions that the stochastic background is
\aref{gaussian} Gaussian,
\aref{ergodic} ergodic,
and
\aref{stationary} stationary,
with no correlation between amplitudes from different 
\aref{indepsky} sky locations or 
\aref{indeppol} polarizations, and with 
\aref{equipartition} equipartition of power across polarizations; and also, commonly (although not universally) assumed to be 
\aref{isotropic} isotropic.
While we find that the first four of these premises are generally applicable beyond GR, the same is not true for the rest---this is without considering changes to the potential sources of the background in beyond-GR theories, which may themselves break more of the assumed symmetries.
In particular, it is not reasonable to always assume that the usual linear GW polarization amplitudes of \eq{polarizations} will be statistically independent and have well-defined phase velocities, as this will not be true unless the chosen polarization basis diagonalizes the kinetic matrix of the underlying theory of gravity.
Similarly, one should be careful in assuming that power will be equipartitioned among polarizations, even for modes with the same spin-weight, as parity-asymmetric theories may predict differences in the generation and propagation of modes with different helicities.
Deviations from isotropy should be expected in theories with intrinsically preferred frames.

Finally, we have provided specific examples of beyond-GR theories in which these traditional assumptions break down, and in which the GR expression for the stress-energy of a gravitational wave may receive a correction: Chern-Simons gravity, scalar-tensor theories, and massive gravity.
For all these theories, we find that the cross-correlation of the outputs of two ideal differential-arm detectors can be written in terms of the effective GW stress-energy as in \eq{result}, with different $\Xi(f)$ factors encoding how each theory departs from GR.
This set of examples is not intended to be exhaustive, but merely to show that it is possible to construct viable theories that violate standard assumptions in stochastic searches. 
This will be important in the interpretation of results like \cite{Callister2017, nongrsgwo1} once a stochastic signal is detected.

\begin{acknowledgments}

The authors would like to thank
Laura Bernard,
Tom Callister,
Claudia de Rham,
Kurt Hinterbichler,
Andrew Matas,
and
Andrew Tolley
for useful discussions.
M.I.~is a member of the LIGO Laboratory.
LIGO was constructed by the California Institute of Technology and Massachusetts Institute of Technology with funding from the National Science Foundation and operates under cooperative agreement PHY--0757058.
L.C.S.~acknowledges the support of NSF grant PHY--1404569, and the Brinson Foundation.
This paper carries LIGO Document Number \dcc.
\end{acknowledgments}

\appendix
\section{Plane-wave decomposition} \label{app:planewave}

Begin with our compact expression for the plane-wave expansion of the metric components, \eq{h_expansion}:
\beq \label{eq:h_expansion_app}
h_{ab}(\fvec{x}) = \frac{1}{2\pi} \int \hf_{ab}(\fvec{k}) e^{i\fvec{k} \cdot
\fvec{x}}~\dk\, ,
\eeq
with the integral over the four-wave-vector $\fvec{k}$ as prescribed by our definition of $\dk$ in \eq{dk},
\beq  \label{eq:dk_app}
\dk \equiv 2 c\, \delta(|\vec{k}|^2 - |\vec{k}_\omega|^2)\, |\vec{k}|^{-1} \infd
\fvec{k}\, .
\eeq
This definition of the four-dimensional Fourier transform is designed to yield \eq{h_expansion_long}, and thus follows the convention of recent stochastic GW background literature (e.g.\ \cite{Flanagan1993, Allen1999, Romano2017}).
This choice, however, differs from the Lorentz-invariant measure most
common in field theory (see e.g.\ Eq.\ (3.18) in \cite{SrednickiQFT}
or Eq.\ (4.4) in \cite{RyderQFT}),
\begin{align}
  \dk_{\rm QFT} & = c \delta(|\vec{k}|^2 - |\vec{k}_\omega|^2)\, \infd \fvec{k}/(2\pi)^3\, \nonumber \\
  &= \normk \dk /2/(2\pi)^{3} \,.
\end{align}
Note that this difference in measures results in a difference in conventions for the Fourier amplitudes.
Specifically, this means that $\hf(k) \propto \normk \hf(k)_{\rm QFT}  $ (the factor of proportionality depends on prefactors outside of the integral).

With the help of \eq{dk_app}, \eq{h_expansion_app} can be immediately
rewritten as an explicit integral over the four-vector $\fvec{k}$,
transforming each component independently,
\beq
h_{ab}(\fvec{x}) = \frac{c}{\pi} \int \hf_{ab}(\fvec{k}) e^{i\fvec{k} \cdot \fvec{x}}~ \delta(|\vec{k}|^2 - |\vec{k}_\omega|^2)\, \normk^{-1} \infd \fvec{k}\, .
\eeq
Here $\vec{k}_\omega \equiv \vec{k} (\omega)$ encodes the functional dependence of $\vec{k}$ on $\omega$ imposed by the specific dispersion relation required by the underlying theory of gravity---in GR, this is just the usual demand that $|\vec{k}_\omega| = \omega/c$).
For clarity, we may split the 4-vector $\fvec{k}$ into frequency and
spatial $\vec{k}$-vector,
\begin{align}
 h_{ab}(t, \vec{x}) &= \frac{1}{\pi} \int_{-\infty}^\infty \int_{S^2} \int_{0}^\infty \hf_{ab}(\omega, \vec{k})\,
 e^{i(\vec{k}\cdot\vec{x} - \omega t) }~ \nonumber \\
&\times \delta(|\vec{k}|^2 - |\vec{k}_\omega|^2)\, \normk\, \infd \normk \, \infd \hat{k} \, \infd \omega \, ,
\end{align}
where we have written the spatial three-integral in polar coordinates such that 
\beq
\infd \vec{k} = \normk^2\, \infd\normk \, \infd \hat{k}\, ,
\eeq
with angular domain over the 2-sphere, $S^2$.
In this step, we have also used the fact that $\normk$ is non-negative by definition to set its integration limits.

We may now use the fact that, for any continuously differentiable function $g(x)$ with real roots $x_i$,
\beq
\delta( g(x) ) = \sum_i \frac{\delta(x - x_i)}{|g'(x_i)|}\, ,
\eeq
to further simplify the integrand to
\begin{align}
 h_{ab}(t, \vec{x}) &= \frac{1}{2\pi} \int_{-\infty}^\infty \int_{S^2} \int_{0}^\infty \hf_{ab}(\omega, \vec{k})\,
 e^{i(\vec{k}\cdot\vec{x} - \omega t) }~ \nonumber \\
&\times \frac{1}{|\vec{k}_\omega|} \delta(|\vec{k}| - |\vec{k}_\omega|)\, \normk\, \infd \normk \, \infd \hat{k} \, \infd \omega \, ,
\end{align}
where the integration limits have allowed us to ignore the negative root, $\normk = - |\vec{k}_\omega|$.
It is now straightforward to carry out the integral over the norm $\normk$ to obtain:
\beq
h_{ab}(t, \vec{x}) = \frac{1}{2\pi} \int_{-\infty}^\infty \int_{S^2} \hf_{ab}(\omega, \vec{k})
e^{i(\vec{k}\cdot\vec{x} - \omega t) }
\infd \hat{k} \infd \omega\, ,
\eeq
where now $\vec{k}$ is necessarily on shell ($\normk = |\vec{k}_\omega|$).
Writing this in terms of $f=\omega/2\pi$, $\dir = -\hat{k}$ and $v_{\rm p}\equiv|\vec{k}/\omega|^{-1}$, we immediately recover \eq{h_expansion_long}, as promised,
\beq
h_{ab}(t, \vec{x}) = \int_{-\infty}^\infty \int_{\rm sky} \hf_{ab}(k, \dir)
e^{-2\pi i f (t + \dir\cdot\vec{x}/ v_{\rm p}) } \infd \dir \infd f\, ,
\eeq
thus justifying the second equality in \eq{dk},
\beq
\dk = \infd \omega \, \infd \dir\, .
\eeq

\section{Correlation and spectral density} \label{app:correlation}

We will reproduce the standard result that assumptions \aref{stationary} of stationarity and \aref{indepsky} of uncorrelated sky locations allow us to write the cross-correlation of the Fourier amplitudes as in \eq{correlation},
\beq \label{eq:correlation_app}
\left\langle \hf^*_A(\fvec{k})\, \hf_{A'}(\fvec{k}') \right\rangle = \frac{1}{2} \delta(f-f')\, \delta(\dir - \dir')\, \PSDsky_{AA'}(\fvec{k})\, , 
\eeq
with $\PSDsky_{\lp\lp'}$ the cross-power spectral density of stochastic signals of polarizations $\lp$ and $\lp'$.

The second delta function in \eq{correlation_app} is just a direct statement of assumption \aref{indepsky}, so focus on the rest of the equation by suppressing the dependence on $\dir$.
We are then left with simple one-dimensional Fourier transforms in the expression for the cross-correlation,
\begin{align}
\left\langle \hf^*_A(f) \hf_{A'}(f') \right\rangle = 
&\left\langle \int h_A(t) e^{2\pi i f t} \infd t \right. \\ 
\times &\left. \int h_{A'}(t') e^{-2\pi i f' t'} \infd t'\right\rangle .\nonumber
\end{align}
Defining $\tau\equiv t' - t$, this can be put in the form:
\begin{align}
\left\langle \hf^*_A(f)\hf_{A'}(f') \right\rangle =
\int &\int \left\langle h_A(t) h_{A'}(t+\tau)\right\rangle \\
&\times e^{-2\pi i f' \tau} e^{2\pi i (f-f')t} \infd \tau \infd t \, .\nonumber
\end{align}
Now note that the term in brackets is simply the correlation of $h_A(t)$ and $h_{A'}(t')$, which by assumption of stationarity depends only on the time difference $\tau$, i.e.
\beq
\left\langle h_A(t) h_{A'}(t+\tau)\right\rangle =
\left\langle h_A(0) h_{A'}(\tau)\right\rangle\, ,
\eeq
where we have set $t=0$ for concreteness.
We may therefore carry out the integral over $t$ to obtain
\beq
\left\langle \hf^*_A(f)\hf_{A'}(f') \right\rangle = \delta(f-f')\int \left\langle h_A(0) h_{A'}(\tau)\right\rangle e^{-2\pi i f \tau} \infd \tau \, .
\eeq
Now, the Wiener-Khinchin theorem~\cite{Jaranowski2009} guarantees that, if the cross-correlation is continuous, we can always define a function of frequency to give the Fourier transform of the cross-correlation (the integral above); that can be taken as the {\em definition} of the one-sided cross-power spectral density,
\beq
\PSD_{XY} (f) \equiv 2 \int \left\langle X(0) Y(\tau)\right\rangle e^{-2\pi i f \tau} \infd \tau\, ,
\eeq
for any two stationary random processes, $X(t)$ and $Y(t)$, and where the prefactor is chosen so that $\PSD(f)\equiv\PSD(|f|)$ is the {\em one-sided} spectral density.
All this means is that we may write
\beq
\left\langle \hf^*_A(f)\, \hf_{A'}(f') \right\rangle = \frac{1}{2} \delta(f-f') \PSD_{AA'} (f)\, ,
\eeq
or, restoring the $\dir$ dependence,
\beq
\left\langle \hf^*_A(\fvec{k})\, \hf_{A'}(\fvec{k}') \right\rangle = \frac{1}{2} \delta(f-f') \delta(\dir-\dir') \PSDsky_{AA'} (\fvec{k})\, .
\eeq

\section{Scalar-tensor computations} \label{app:st}

Here we provide details on the computations of the ESET and correlation functions in Brans-Dicke gravity (\sect{st}).
In order to do so, first consider the transformations between the Jordan and Einstein frames.
By definition of the Einstein frame, in a generic scalar-tensor theory these can be written as (e.g.\ Eqs.\ (34)--(36) in \cite{Will2006})
\begin{subequations} \label{eq:st_transf_app}
\begin{align}
\gstj_{ab} &\equiv A^{2}(\phie) \gste_{ab} \, ,
\\
\phij &\equiv A^{-2}(\phie)\, ,
\end{align}
\end{subequations}
for some auxiliary function $A(\phie)$.
We can then use this to define the coupling $\alpha(\phie)$ as
\beq
\alpha (\phie) \equiv \frac{\infd \ln A(\phie)}{\infd \phie}\, .
\eeq
To recover the Brans-Dicke theory, we simply expand this coupling to linear order by setting
\beq \label{eq:alpha0_app}
\alpha(\phie) = \alpha_0 \equiv (2\obd + 3)^{-1/2}\, ,
\eeq
so that $\ln A(\phie) = \alpha_0 (\phie -{\phie}_0)$ for some fiducial value $\phie_0$, and Eqs.\ \eqref{eq:st_transf_app} become (Jordan to Einstein)
\begin{subequations} \label{eq:bd_transf}
\begin{align}
\gstj_{ab} &= e^{2\alpha_0 (\phie-\phie_0)} \gste_{ab}\, ,
\\
\phij &= e^{-2\alpha_0 (\phie - \phie_0)}\, .
\end{align}
\end{subequations}

For later convenience, define $\delta\phie \equiv \phie - \phie_0$ and rescale the Jordan field by letting $\phij \rightarrow \phij/\phij_0$ for some background value $\phij_0$.
After doing so, Eqs.\ \eqref{eq:bd_transf} imply (Einstein to Jordan)
\begin{subequations} \label{eq:bd_transf_3}
\begin{align}
\gste_{ab} &= \frac{\phij}{\phij_0} \gstj_{ab}\, ,
\\
\delta \phie &= - \frac{\ln (\phij/\phij_0)}{2\alpha_0}\, .
\end{align}
\end{subequations}

With the above notation in place, let us perturb the two metrics and scalar fields to first order, and then obtain the relationship between the perturbations in the two frames.
Letting
$\gstj_{ab} \rightarrow \gstj_{ab} + \hste_{ab}$ and $\phij \rightarrow \phij_0 + \delta \phij$
in the Jordan frame, and $\gste_{ab} \rightarrow \gste_{ab} + \hste_{ab}$ and $\phie \rightarrow \phie_0 + \delta \phie$ in the Einstein frame, we can then apply the transformations from \eq{bd_transf_3} to write
\begin{align}
\gste_{ab} + \hste_{ab} &= \phij_0^{-1} \left(\phij_0 + \delta \phij \right)\left(\gstj_{ab} + \hstj_{ab}\right) \nonumber \\
&\approx \gstj_{ab} + \left(\hstj_{ab} + g_{ab} \delta\phij/\phij_0\right) .
\end{align}
Collecting terms of the same order, this implies that, to first order in the perturbations,
\begin{subequations}
\begin{align}
\gstj_{ab} &= \gste_{ab}\, ,
\\
\hstj_{ab} &= \hste_{ab} + 2\alpha_0 \delta\phie \gste_{ab}\, ,
\\ \label{eq:Phi_delta-phie}
\Phi &= 2\alpha_0 \delta\phie\, ,
\end{align}
\end{subequations}
where we have defined $\Phi \equiv - \delta\phij/\phij_0$ for convenience.
Using this definition to replace the second expression by $\hste_{ab} = \hstj_{ab} - \Phi \gstj_{ab}$, it becomes clear these are Eqs.\ \eqref{eq:st_lin-transf} provided in the main text.

\subsection{Effective stress-energy tensor} \label{app:st_eset}

\newcommand{\htre}{\ste{\htr}}

We wish to compute the effective GW stress energy in the Einstein frame.
We will do so by taking advantage of the gauge proposed in \cite{Lee:1974pt}, in which the trace-reversed Einstein-frame perturbation, $\ste{\htr}_{ab}$, satisfies
\begin{subequations} \label{eq:ste_constr}
\begin{align}
\label{eq:htre_trace}
\htre = 2 \Phi\, ,
\\ \label{eq:htre_div}
\nabla^a \htre_{ab} = 0\, ,
\end{align}
\end{subequations}
and follows simple free-wave equations of motion,
\begin{subequations} \label{eq:ste_eom}
\begin{align}
\label{eq:htre_box}
\Box \htre_{ab} &= 0\, ,
\\ \label{eq:phie_box}
\Box \Phi &= 0\, .
\end{align}
\end{subequations}
In this gauge, the Einstein-frame trace-reversed metric perturbation is equal to the regular (non-trace-reversed) perturbation in the Jordan frame: $\ste{\htr}_{ab} = \hstj_{ab}$.
Thus, $\ste{\htr}_{ab}$ may be decomposed into synchronous polarizations as in \eq{st_hpols}.

To obtain an expression for the GW stress energy in the Einstein frame, we may follow the procedure outlined in \sect{energy} starting from the action of \eq{sste}. 
Perturbing the metric and scalar as described above, and discarding terms higher than second order, we may obtain the quadratic Lagrangian density corresponding to $\act^{(2)}$ in \eq{spert},
\beq \label{eq:ste_L2}
\ste{{\cal L}}^{(2)} =  \ste{{\cal L}}_{\rm EH}^{(2)} + \kappa \sqrt{-\gste} \left[ - 2 \gste^{ab} \nabla_a ( \delta \phie) \, \nabla_b (\delta \phie) \right] ,
\eeq
where $\ste{{\cal L}}_{\rm EH}^{(2)}$ is the Einstein-Hilbert piece of
\eq{lmt}, but in terms of $\gste, \hste$.
The variation of this quantity with respect to $\ste{\htr}^{ab}$ and $\delta\phie$ will lead to the ESET per \eq{tgw}.
This will be given by a contribution from the Einstein-Hilbert part of the action (the Ricci terms above), and another from the rest.
We will call those two terms $\ste{T}^{\rm (EH)}_{ab}$ and $\Delta \ste{T}^{\rm (ST)}_{ab}$ respectively, so that $\ste{T}_{ab}^{\rm (ST)} = \ste{T}^{\rm (EH)}_{ab} + \Delta \ste{T}^{\rm (ST)}_{ab}$.

Focus first on the EH term.
This will {\em not} be the identical to \eq{isaacson} in GR, because there will be an extra contribution from the nonvanishing trace of $\htre_{ab}$, \eq{htre_trace}.
To compute it, we may take a shortcut and begin from an expression obtained MacCallum and Taub for the effective EH quadratic Lagrangian contributing to the GW stress-energy far away from the source \cite{MacCallum1973}.
The corresponding stress-energy tensor can be written as
\begin{align} \label{eq:tmt}
\ste{T}^{\rm (EH)}_{ab} = \kappa &\llangle[\Bigg]
\frac{1}{2} \nabla_a \htre^{cd} \nabla_b \htre_{cd}
- \nabla_c \htre_{da} \nabla^c \htre_b{}^d
-\frac{1}{4} \nabla_a \htre \nabla_b \htre \nonumber \\
&+ \frac{1}{2} \nabla_c \htre_{ab} \nabla^c \htre
+ \gste_{ab} \left( \frac{1}{2} \nabla_e \htre^{fc} \nabla_f \htre^e{}_c \right. \nonumber \\
&\left. - \frac{1}{4} \nabla_e \htre_{cd} \nabla^e \htre^{cd}
+ \frac{1}{8} \nabla_e \htre \nabla^e \htre \right)\rrangle[\Bigg] \, .
\end{align}
This expression is valid whenever separation of length-scales allows for a clear definition of the waves over some background.
In GR, application of the equations of motion in a
transverse-traceless gauge reduces \eq{tmt} to the Isaacson formula,
\eq{isaacson}.  We proceed similarly here but keeping the trace, using
Eqs.\ \eqref{eq:ste_eom} and \eqref{eq:ste_constr}.

First note that the second term in \eq{tmt} may be re-written by
integrating by parts ``under the average.''  This is because the
Brill-Hartle average of a total derivative is smaller by a factor of
order ${\cal O}(\lambda_{\rm GW}/L_{\rm ave})$ than non-vanishing
averages, where $L_{\rm ave}$ is the averaging length scale (see e.g.\
Sect.\ IIA in \cite{Stein2011}).  This then implies that
\beq
\label{eq:int-by-parts-ex}
\llangle - \nabla_a \htre_{bc} \nabla^a \htre_d{}^b\rrangle = \llangle \left(\nabla^a \nabla_a \htre_{bc} \right) \htre_d{}^b \rrangle 
\left[1 + {\cal O}\hspace{-2pt}\left(\frac{\lambda_{\rm GW}}{L_{\rm ave}}\right)\right].
\eeq
Therefore, the second term in \eq{tmt} vanishes via the equations of
motion [\eq{htre_box}], up to this order.
The same logic may be applied to all terms in the second and third lines of \eq{tmt}, which will vanish due to \eq{htre_box} or \eq{htre_div}.
We are then only left with the first and third terms in \eq{tmt}.
The first term is just the same quadratic contribution that appears in \eq{isaacson} for GR.
Meanwhile, the third term involves the trace of $\htre_{ab}$, and may thus be written in terms of the scalar field using \eq{htre_trace}.
The contribution of the Einstein-Hilbert part of the action to the ESET, \eq{tmt}, in ST gravity then reduces to
\beq
\ste{T}^{\rm (EH)}_{ab} = \frac{1}{2} \kappa \llangle \nabla_a \htre^{cd} \nabla_b \htre_{cd} \rrangle - \kappa \llangle \nabla_a \Phi \nabla_b \Phi \rrangle \, .
\eeq

Switch now to the contribution from the kinetic term of the scalar field, $\Delta\ste{T}_{ab}^{\rm (ST)}$.
This will be obtained from the corresponding part of the quadratic Lagrangian of \eq{ste_L2}, namely $\Delta \ste{\cal L}^{(2)} \equiv -2 \sqrt{-\ste{g}} \kappa \gste^{ab} \nabla_a (\delta \phie) \nabla_b (\delta \phie)$.
The variation of this quantity may be written as
\beq
\frac{\delta \Delta \ste{\cal L}^{(2)}}{\delta \gste^{ab}} =
\kappa \sqrt{-\gste} \left[ \gste_{ab} \gste^{cd}
- \delta^c{}_{(a} \delta^d{}_{b)} \right]
\nabla_c \delta \phie\, \nabla_d \delta \phie \, ,
\eeq
using the usual fact that $\delta \sqrt{-\gste} = - \sqrt{-\gste}\, \gste_{ab} \delta\gste^{ab}/2$, and explicitly symmetrizing the variation of the metric.
Therefore, \eq{tgw} implies that
\begin{align}
\Delta \ste{T}_{ab}^{\rm (ST)}
&= 2 \kappa \llangle[\Big] \left(- \gste_{ab} \gste^{\alpha\beta}
+ 2 \delta^\alpha{}_{a} \delta^\beta{}_{b} \right)
\nabla_\alpha \delta \phie\, \nabla_\beta \delta \phie \rrangle[\Big] \nonumber \\
&= 4 \kappa \llangle \nabla_a \delta \phie\, \nabla_b \delta \phie \rrangle \, , 
\end{align}
where one the first term vanished due to \eq{phie_box}, by integration by parts under averaging as before.

We may now write an expression for the total effective stress energy of a scalar-tensor GW in the Einstein frame:
\begin{align}
\ste{T}^{\rm (ST)}_{ab} &= \ste{T}_{ab}^{\rm (EH)} + \Delta \ste{T}_{ab}^{\rm (ST)} \\
&= \kappa \llangle[\Bigg] \frac{1}{2} \nabla_a \htre^{cd} \nabla_b \htre_{cd}
+ \left(\alpha_0^{-2} -1\right) \nabla_a \Phi \nabla_b \Phi  \rrangle[\Bigg] , \nonumber
\end{align}
where we have used the fact that $\delta \phie = \Phi/(2\alpha_0)$ to first order, \eq{Phi_delta-phie}.
This may also be written in terms of the Brans-Dicke parameter using the definition of $\alpha_0$, \eq{alpha0_app}, to obtain our final result presented in \eq{tst_all}.

\subsection{Energy density spectrum} \label{app:st_rho}

Taking the time-time component of \eq{tst_all} and assuming \aref{ergodic} ergodicity, we immediately obtain an expression for $\rgw$ in a local Lorentz frame ($\ste{g}_{ab}=\eta_{ab}$) from \eq{tst_all},
\beq \label{eq:st_app_rgw_hij}
\rgw  \hspace{-1.5pt} = \hspace{-1.5pt} \frac{\kappa}{2c^2} \left[ \left\langle \partial_t \htre^{ij} \partial_t \htre_{ij} \right\rangle \hspace{-1pt}
+ \hspace{-1pt} 4 (\obd + 1 ) \hspace{-1.5pt} \left\langle \partial_t \Phi \partial_t \Phi \right\rangle \right] \hspace{-2pt}, 
\eeq
or equivalently, because $\Phi = \ste{g}^{ab} \ste{h}_{ab}/2$ by \eq{st_hpols},
\begin{align}
\rgw = \frac{\kappa}{2c^2} &\left[ g^{ik} g^{jl} + \left(\obd + 1 \right) g^{ij} g^{kl} \right] \nonumber\\
&\times \left\langle \partial_t \htre_{ij} \partial_t \htre_{kl} \right\rangle .
\end{align}
Expanding out $\htre_{ij}$ into plane-waves, $\rgw$ can then be put in the form of \eq{rgw-k-space-double-integ} with ${\cal Q}^{abcd}$ as in \eq{Qst} of the main text. 

For convenience, denote each of the two terms in \eq{st_app_rgw_hij} $\rho_{\rm EH}$ and $\rho_{\rm ST}$ respectively, so that $\rgw = \rho_{\rm EH} + \rho_{\rm ST}$.
Making use of all the usual assumptions \aref{gaussian}--\aref{isotropic} about the background, we can use \eq{correlation_isotropic} to write $\rho_{\rm EH}$ in the same form as \eq{rgw_gr} in GR,
\beq
\rho_{\rm EH} = \frac{\pi c^2}{4G} \sum_{\lp} \int_0^\infty S_{\lp} (f) f^2 \,  \infd f \, ,
\eeq
except that now the sum is over $\lp \in \{+, \times, {\rm b}\}$ with $S_{\rm b} = S_\Phi$ by \eq{st_hpols}.
For $\rho_{\rm ST}$, a similar derivation to the one for $\rho_{\rm EH}$ gives the analogous result that
\beq
\rho_{\rm ST} = \frac{\pi c^2}{4G} \left(2 + 2\obd\right) \int_0^\infty S_{\rm b} (f) f^2\, \infd f\, .
\eeq
Adding both contributions together, we may then write the total energy spectrum compactly
as we did in Eqs.\ \eqref{eq:stfactor} and \eqref{eq:strgw} in the main text.

\section{Massive gravity computations} \label{app:mg}

Here we provide more details for the computation of the GW stress-energy density and correlation functions presented in \sect{mg}.
In App.\ \ref{app:mg_pols} we derive the expressions for the unitary and synchronous metric components, presented respectively in Eqs.\ \eqref{eq:h_mg_unitary} and \eqref{eq:h_mg_transf} in the main text.
In App.\ \ref{app:mg_energy}, we obtain an expression for the ESET in Fierz-Pauli massive gravity, and one for the energy density $\rgw$ in terms of the unitary PSDs, making use of statistical assumptions about the background.
Finally, in App.\ \ref{app:mg_crosscor} we compute an expression for the cross-correlation of the output of two differential-arm detectors in the form of \eq{result}.
We will make repeated use of the massive Klein-Gordon equation of motion of \eq{mg_box}, as well as the fact that the metric perturbation must be divergenceless, \eq{mg_div}, and traceless, \eq{mg_trace}.
Throughout this appendix, ``massive gravity'' refers to the Fierz-Pauli theory of \eq{spf}.

\subsection{Polarizations} \label{app:mg_pols}

\subsubsection{Unitary gauge} \label{app:mg_unitary}

We would like to decompose a massive plane GW into a basis of
polarization tensors.
In GR, diffeomorphism invariance guarantees that we may always find a gauge in which the perturbation is purely spatial, as in \eq{polarizations}.
Although this is not possible in FP gravity, we may still write a generic metric perturbation propagating in the $z$-direction, $h_{\mu\nu}$, as
\beq \label{eq:mg_hu_ansatz}
(h_{\mu\nu}) = \begin{pmatrix}
h_{00} & h_{01} & h_{02}    & h_{03}  \\
h_{10} & h_{\rm b} + h_+ & h_\times    & h_{\rm x}  \\
h_{20} & h_\times    & h_{\rm b} - h_+ & h_{\rm y}  \\
h_{30} & h_{\rm x}    & h_{\rm y}    & h_{\rm l}
\end{pmatrix} ,
\eeq
and then apply the constraints from Eqs.\ \eqref{eq:mg_constraints} to cut down the number of degrees of freedom.

First, for a GW with wave-vector $k^a$, \eq{mg_div} implies $k^a h_{ab}=0$.
Thus, picking a frame in which the wave travels in the $z$-direction,
\beq \label{eq:mg_k}
(k^\mu) = \left(\omega/c,\, 0,\, 0,\, \normk\right) ,
\eeq
lack of divergence, together with symmetry, must mean
\beq \label{eq:mg_div_condition}
h_{0\mu} = h_{\mu 0} = -\mgnullity h_{3\mu} = - \mgnullity h_{\mu3}\, ,
\eeq
with $\mgnullity$ as in \eq{mg_ratio_2}.
Equation~\eqref{eq:mg_div_condition} also implies that
\beq
h_{00} = - \mgnullity h_{30} = -\mgnullity h_{03} = \mgnullity^2 h_{33}\, .
\eeq
However, tracelessness, \eq{mg_trace}, also demands
\beq
h = \eta^{\mu\nu} h_{\mu\nu} = -h_{00} + 2 h_{\rm b} + h_{\rm l} = 0\, .
\eeq
Therefore, if we choose to get rid of the time-time degree of freedom by writing $h_{00} = 2 h_{\rm b} + h_{\rm l}$, \eq{mg_constraints} requires
\beq \label{eq:mg_trace_condition}
2 h_{\rm b} + h_{\rm l} =  \mgnullity^2 h_{\rm l} \implies h_{\rm b} = \frac{1}{2}\left(\mgnullity^2-1\right) h_{\rm l}\, ,
\eeq
so we will only need one scalar polarization.
This could have been anticipated from the fact that a symmetric rank-2 tensor in four dimensions can have at most ten independent components, five of which are necessarily constrained by \eq{mg_constraints}, leaving only five degrees of freedom.
These correspond to the five possible helicities of a massive spin-2 particle.

Choosing to work in terms of the longitudinal scalar amplitude, our final expression for the metric perturbation in the {\em unitary gauge} native to FP gravity is \eq{h_mg_unitary}, i.e.
\beq \label{eq:h_mg_unitary_app}
(h_{\mu\nu}) = \begin{pmatrix}
\mgnullity^2 h_{\rm l}  & -\mgnullity h_{\rm x} & -\mgnullity h_{\rm y}    & -\mgnullity h_{\rm l}  \\
- \mgnullity h_{\rm x} & -\frac{1}{2}\mgratio^2 h_{\rm l} + h_+ & h_\times    & h_{\rm x}  \\
- \mgnullity h_{\rm y} & h_\times    & -\frac{1}{2}\mgratio^2 h_{\rm l} - h_+ & h_{\rm y}  \\
- \mgnullity h_{\rm l} & h_{\rm x}    & h_{\rm y}    & h_{\rm l}
\end{pmatrix} ,
\eeq
where we have used the fact that $(\mgnullity^2-1)=\mgratio^2$ by \eq{mg_ratio_2}.

For later convenience, note that the metric perturbation of \eq{h_mg_unitary_app} satisfies
\beq \label{eq:mg_id1}
h^{ab}(\fvec{k}) h_{ab}'(\fvec{k}') = 2 \sum_{\lp} C_{\lp}(\omega, \omega')\thinspace h_{\lp}(\fvec{k})\thinspace h_{\lp}'(\fvec{k}')\, ,
\eeq
with a sum over unitary polarizations $\lp$, and for some frequency-dependent normalization coefficients $C_{\lp}$ defined by
\beq \label{eq:mg_id1_coeff}
C_\lp(\omega, \omega') \equiv \begin{cases} 
1 & {\rm if~} \lp = +,\times , \\
1 - \mgnullity\mgnullity' & {\rm if~} \lp ={\rm x},{\rm y} ,\\
\frac{3}{2}\left(1-\mgnullity\mgnullity'\right)^2 -\frac{1}{2}\left(\mgnullity - \mgnullity'\right)^2 & {\rm if~} \lp = {\rm l}\, .
\end{cases}
\eeq
The form of these coefficients should not come as a surprise, since they are just terms of the form $e^\lp{}_{ab} e_\lp{}^{ab}$, analogous to those in \eq{lp_norm}, times extra factors arising from the trace and timelike components of \eq{h_mg_unitary_app}.
Note that $C_{\lp}$ acquires its frequency dependence via $\mgnullity$ and $\mgnullity'$, \eq{mg_ratio_2}.

\subsubsection{Synchronous gauge} \label{app:mg_sync}

\newcommand{\hvx}{h_{\rm x}}
\newcommand{\hvy}{h_{\rm y}}
\newcommand{\hsl}{h_{\rm l}}

As mentioned in Sec.~\ref{sec:detection}, it is easiest to compute the
influence of a gravitational wave on a LIGO-style detector in the
synchronous gauge, because the coordinate locations of the
mirrors do not change in this gauge~\cite{MTW}.  However, massive gravity lacks
the linearized diffeomorphism freedom needed to transform into
synchronous gauge.
Fortunately, we circumvent the lack of linearized diffeomorphism
invariance in massive gravity by using the {\em \Stuck{} trick}: we
can add extra auxiliary fields to write the FP action, \eq{spf}, as a
gauge-fixed version of a gauge invariant theory \cite{Stueckelberg:1957zz}.
After adding the \Stuck{} fields, $\xi_a$, we will have the usual
freedom to carry out infinitesimal coordinate transformations,
\beq \label{eq:mg_stuck}
\syn{h}_{ab} = h_{ab} + \nabla_a \xi_{b} + \nabla_b \xi_a\, ,
\eeq
We will want to choose the fields $\xi_a$ such that we can go from the unitary gauge $h_{ab}$ of \eq{h_mg_unitary_app} to a synchronous gauge $\syn{h}_{ab}$ in which $\syn{h}_{0\nu}=0$.
To do this, pick the same frame as before, in which $\fvec{k}$ is given by \eq{mg_k}, and use linearity to consider the transformation of the degrees of freedom in \eq{h_mg_unitary_app}, $\{h_+, h_\times, \hvx, \hvy, h_{\rm l} \}$, one by one.
Below, we will temporarily let $c=1$ and $k\equiv\normk$ for simplicity, but the final result of \eq{h_mg_synchronous_app} is insensitive to this.
For simplicity, we also let $g_{ab} = \eta_{ab}$.
(For more details on the application of this technique to massive gravity, see e.g.\ Sect.\ 2.2.2 in \cite{DeRham2014} or Sect.\ IV in \cite{Hinterbichler2011}.)

Because the two tensor degrees of freedom, $h_+$ and $h_\times$, only
appear in the spatial part of \eq{h_mg_unitary_app}, these modes
already satisfy the synchronous gauge condition.
Next consider the vector-x amplitude, $h_{\rm x}$:
to determine the transformation that would make its contributions to time-like components in \eq{h_mg_unitary_app} vanish, suppose the unitary perturbation is given simply by
\beq \label{eq:mg_hstuck_b}
\left(h_{\mu\nu}\right) = \begin{pmatrix}
0  & -\mgnullity h_{\rm x} & 0 & 0  \\
- \mgnullity h_{\rm x} & 0 & 0    & h_{\rm x}  \\
0 & 0    & 0 & 0  \\
0& h_{\rm x}    & 0    &  0
\end{pmatrix} ,
\eeq
and let the single degree of freedom be a simple plane-wave, $\hvx = A_{x} \sin (\omega t - k z)$.
The goal is to find the form of $\xi_\mu$ in \eq{mg_stuck} that yields $\syn{h}_{0\nu} =0$ in this frame.
For instance, for the time-time component, \eq{mg_stuck} and our requirement that $\syn{h}_{00} = 0$ imply
\beq
\partial_0 \xi_0 = 0 \implies \xi_0 = 0\, .
\eeq
In the last step, we integrated over time and used gauge freedom to pick initial conditions in which $\xi_0 (\vec{x})=0$ for all $\vec{x}$, so that we can ignore the integration constant.
Similarly, using this result for $\xi_0$ and demanding $\syn{h}_{01} = 0$, we can also conclude that \eq{mg_stuck} requires
\beq
h_{01} + \partial_0 \xi_1 + \partial_1 \xi_0 = 0 \implies \partial_t \xi_1 = \mgnullity \hvx\, , 
\eeq
which we can integrate, as we did above, to get
\beq
\xi_1
= -\frac{\mgnullity}{\omega} A_x \cos (\omega t - k z) .
\eeq
Since this is the only nonvanishing component of the \Stuck{} field relevant to the vector-x amplitude, \eq{mg_stuck} implies that
\beq
\syn{h}_{13} = \syn{h}_{31} = \hvx + \partial_z \xi_1 = \left(1-\mgnullity^2\right) \hvx,
\eeq
and $\syn{h}_{\mu\nu}=0$ otherwise, for a unitary metric perturbation whose only non-zero components come from $\hvx$, as we supposed above in \eq{mg_hstuck_b}. 
It can be shown that the same exact argument, applied to $\hvy$ instead of $\hvx$, yields an analogous result,
\beq
\syn{h}_{23} = \syn{h}_{32} = \hvy + \partial_z \xi_1 = \left(1-\mgnullity^2\right) \hvy,
\eeq
if we had started with a unitary metric perturbation whose only non-vanishing degree of freedom was $\hvy$.

The case of the longitudinal amplitude, $\hsl$, is slightly more complicated, but can be handled in the same way.
Suppose the perturbation is given simply by
\beq \label{eq:mg_hstuck_l}
(h_{\mu\nu}) = \begin{pmatrix}
\mgnullity^2 h_{\rm l}  & 0 & 0    & -\mgnullity h_{\rm l}  \\
0 & \frac{1}{2}\left(\mgnullity^2-1\right) h_{\rm l}  & 0    & 0  \\
0 & 0    & \frac{1}{2}\left(\mgnullity^2-1\right) h_{\rm l}  & 0  \\
- \mgnullity h_{\rm l} & 0    & 0    & h_{\rm l}
\end{pmatrix} ,
\eeq
and let $\hsl = A_{l} \sin (\omega t - k z)$, as we did above for $\hvx$ (and, implicitly, $\hvy$).
In this case, the requirement that $\syn{h}_{00} = 0$ implies, via \eq{mg_stuck}, that
\beq
h_{00} + 2 \partial_0 \xi_0 = 0 \implies \partial_0 \xi_0 = -\frac{1}{2} \mgnullity^2 \hsl\, , 
\eeq
and so, integrating over time, we conclude that
\beq
\xi_0
= \frac{\mgnullity^2}{2 \omega} A_l \cos (\omega t - k z)\, ,
\eeq
where we have neglected integration constants, as before.
Now, the result for $\xi_0$ and the requirement that $\syn{h}_{03}=0$ mean that \eq{mg_stuck} also implies
\beq
h_{03} + \partial_0 \xi_3 + \partial_3 \xi_0 = 0 \implies \partial_0 \xi_3 = \left(\mgnullity - \frac{\mgnullity^3}{2}\right) \hsl\, ,
\eeq
and so, integrating over time,
\beq
\xi_3 = \frac{\mgnullity}{\omega}\left(\frac{\mgnullity^2}{2}-1\right) A_l \cos (\omega t -k z)\, .
\eeq
Since $\xi_0$ and $\xi_3$ are the only nonvanishing components of the \Stuck{} field relevant to the longitudinal amplitude, \eq{mg_stuck} implies that 
\begin{align}
\syn{h}_{11} = h_{22} = h_{11} = h_{22} ={}& \frac{1}{2} \left( \mgnullity^2 -1\right) \hsl\, , \\
\syn{h}_{33} = h_{33} + 2 \partial_z \xi_3 ={}& \left(\mgnullity^2 - 1 \right)^2 \hsl \, ,
\end{align}
for a unitary metric whose only non-zero components come from $\hsl$, as we supposed above in \eq{mg_hstuck_l}. 

Putting back all degrees of freedom together, we obtain our final expression for the metric perturbation in a synchronous gauge,
\beq \label{eq:h_mg_synchronous_app}
(\syn{h}_{\mu\nu}) = \begin{pmatrix}
0 & 0 & 0 & 0 \\
0 & h_+ -\frac{1}{2} \mgratio^2 h_{\rm l} & h_\times    &  \mgratio^2\hvx  \\
0 & h_\times    & -h_+ - \frac{1}{2}\mgratio^2 h_{\rm l} & \mgratio^2\hvy  \\
0 & \mgratio^2\hvx    & \mgratio^2\hvy    & \mgratio^4 h_{\rm l}
\end{pmatrix} ,
\eeq
with $\mgratio$ as in \eq{mg_ratio}.
In the limit of no graviton mass ($\mgratio \rightarrow 0$), we manifestly recover the transverse-traceless expression familiar from GR without the need for further gauge fixing.

Finally, it will be useful to define a transformation matrix to go from unitary to synchronous polarization amplitudes.
The unitary amplitudes are simply the degrees of freedom appearing in \eq{h_mg_unitary_app}, while the synchronous ones are just 
\beq \label{eq:h_mg_synchronous_pols_app}
(\syn{h}_{\mu\nu}) = \begin{pmatrix}
0 & 0 & 0 & 0 \\
0 & \syn{h}_{\rm b} + \syn{h}_+ & \syn{h}_\times    &  \syn{h}_{\rm x}  \\
0 & \syn{h}_\times    & \syn{h}_{\rm b}-\syn{h}_+ & \syn{h}_{\rm y}  \\
0 & \syn{h}_{\rm x}    & \syn{h}_{\rm y}    & \syn{h}_{\rm l}
\end{pmatrix} ,
\eeq
in full analogy to \eq{polarizations}.
Comparing this definition to \eq{h_mg_synchronous_app}, it can be
easily shown that the transformation matrix $M_{\syn{A}}{}^{B}$
satisfying $\syn{h}_A = M_{\syn{A}}{}^{B} h_B$ is given by \eq{h_mg_transf}.

\subsection{Effective stress-energy tensor} \label{app:mg_energy}

We wish to obtain an expression for the ESET of GWs in Fierz-Pauli massive gravity, following the procedure outlined in \sect{energy}.
To do so, begin with the total FP action of \eq{spf}, $\act_{\rm FP} = \act_{\rm EH} + \act_{m}$,
with $\act_{\rm EH}$ the Einstein-Hilbert action of \eq{sgr}, and $\act_{m}$ the contribution from the scalar field given by \eq{sm}.
All computations in this section will be carried out in the {\em unitary gauge} native to FP gravity, \eq{h_mg_unitary}, since those polarization amplitudes are the fundamental degrees of freedom that we can take to be uncorrelated in this theory (since they diagonalize its kinetic matrix). 

We will consider the two terms in the FP action separately.
As in the scalar-tensor case (Appendix \ref{app:st}), we may obtain the contribution from the Einstein-Hilbert part by starting from the MacCallum--Taub expression for the stress energy, \eq{tmt}.
Unlike for scalar-tensor, however, we may now ignore all terms showing the trace and let $\htr_{ab}=h_{ab}$, thanks to \eq{mg_trace}.
With these simplifications, \eq{tmt} becomes
\begin{align} \label{eq:tmt_notrace}
T^{\rm (EH)}_{ab} = \kappa &\llangle[\Bigg]
\frac{1}{2} \nabla_a h^{cd} \nabla_b h_{cd}
- \nabla_c h_{da} \nabla^c h_b{}^d
 \\
&{} + g_{ab} \left( \frac{1}{2} \nabla_e h^{fc} \nabla_f h^e{}_c - \frac{1}{4} \nabla_e h_{cd} \nabla^e h^{cd}
\right)\rrangle[\Bigg] . \nonumber 
\end{align}
The first term in this expression yields the Isaacson tensor obtained in GR, \eq{isaacson}, except now the sum must include all five polarizations allowed in \eq{h_mg_unitary}, not just the transverse-traceless ones.
The second term may be rewritten via integration by parts ``under the
average,'' as discussed around \eq{int-by-parts-ex},
so that it becomes\footnote{%
The error here arises from the level at which total derivatives
average out to over the length $L_{\rm ave}$.  Naturally this length
needs to be very large compared to the gravitational wavelength, but
its hierarchy with the Compton wavelength is more subtle.  To justify
keeping the $\mu^{2}$ terms, we need the averaging error to be small
compared to the $\mu^{2}$ terms.
}
\beq
\llangle - \nabla_c h_{da} \nabla^c h_b{}^d  \rrangle
= \mG^2 \llangle h_{da} h_b{}^d \rrangle
+ \text{(avg.~error)}
\,,
\eeq
after applying the equations of motion, \eq{mg_box}.
A similar argument shows that the third term vanishes due to \eq{mg_div}, while the fourth and final term takes a similar form as the second one,
\beq
\llangle - \nabla_e h_{cd} \nabla^e h^{cd} \rrangle
= \mG^2 \llangle h^{cd} h_{cd} \rrangle
+ \text{(avg.~error)}
\, .
\eeq
Altogether, this means that the contribution to the ESET from the Einstein-Hilbert part of the action is
\begin{align}
T^{\rm (EH)}_{ab} &= \frac{\kappa}{2} \llangle \nabla_a h^{cd} \nabla_b h_{cd} \rrangle \\
&+ \kappa \mG^2 \llangle h_{da} h_b{}^d
+ \tfrac{1}{4} g_{ab} h^{cd} h_{cd}  \rrangle\, . \nonumber 
\end{align}

Now focus on the contribution from $\act_{m}$ in \eq{sm}.
This action is already the quadratic action $\act^{(2)}$ needed for
\eq{spert}, namely
\beq
{\cal L}_{m}^{(2)} = \frac{1}{4} \kappa \mG^2 h^{ab} h^{cd}
 \sqrt{-g}\left(g_{ab}g_{cd} - g_{ac} g_{bd} \right) ,
\eeq
where we have explicitly written out the antisymmetrized terms.
We have also written $h^{ab}$ with indices up, to match the index
position convention used in~\cite{Stein2011} and thus in deriving
\eq{tmt}.\footnote{%
  This is a somewhat subtle point, since wrong index position
  generates implicit dependence on the (inverse) metric.  Ultimately
  it doesn't matter whether $h_{ab}$ or $h^{ab}$ is treated as the
  fundamental variable, so long as the same choice is made for all
  parts of the action when performing the variation with respect to
  $g^{ab}$.
  }
The variation of this quantity with respect to the inverse metric can be shown to be
\begin{align}
\frac{\delta {\cal L}_{m}^{(2)}}{\delta g^{cd}} =
  \tfrac{1}{2}\kappa \mG^2 \sqrt{-g} &\left[
  h_{ca} h_{d}{}^{a}
  + \tfrac{1}{4} g_{cd} h_{ab}h^{ab} \right] ,
\end{align}
where we have used the fact that
$\delta g_{ab} = - g_{ac}(\delta g^{cd})g_{db}$,
$\delta \sqrt{-g} = - \sqrt{-g}\,  g_{ab} \delta g^{ab}/2$, and that, on shell, the perturbation is traceless by \eq{mg_trace}.
The contribution of $\act_m$ to the stress energy may be obtained directly from this variation using \eq{tgw}:
\begin{align}
\Delta T_{ab}^{\rm (FP)}
&= -\kappa \mG^2  \llangle[\Big] h_{da} h_b{}^d
 +\frac{1}{4}g_{ab} h^{cd} h_{cd} \rrangle[\Big] \,.
\end{align}

Adding both contributions computed above, the total ESET in massive gravity, $T_{ab} =  T^{\rm (EH)}_{ab} + \Delta T^{\rm (FP)}_{ab}$, is then
\begin{align} \label{eq:tfp_app}
T_{ab} &= \frac{\kappa}{2} \llangle \nabla_a h^{cd} \nabla_b h_{cd} \rrangle 
\,,
\end{align}
as presented in Eqs.\ \eqref{eq:tfp_all}--\eqref{eq:tfp_sum}.
We further discuss the interpretation of this result in the main text.

We now compute an expression for $\rgw$ as a function of the PSD of the unitary polarization amplitudes of \eq{h_mg_unitary_app} [\eq{h_mg_unitary} in the main text].
Expanding the metric perturbation into plane waves in the local Lorentz frame of the detector (with $g_{ab}=\eta_{ab}$), as in \eq{h_expansion}, and taking the time-time component of the ESET, we get
\begin{align}
\rgw &\equiv \frac{\kappa}{2c^2} \left\langle \partial_t h^{\alpha\beta} \partial_t h_{\alpha\beta} \right\rangle \\
&= \frac{-\kappa}{2c^2} \frac{1}{4\pi^2} \int \left\langle \hf^*_{\alpha\beta} (-\fvec{k}) \hf^{\alpha\beta}(\fvec{k}')\right\rangle e^{i(\fvec{k}+\fvec{k}')\cdot\fvec{x}} \omega \omega' \dk \dk',
\nonumber
\end{align}
assuming \aref{ergodic} ergodicity as usual.
The second equality was obtained by proceeding identically as in GR (\sect{gr}).
The contraction inside the angular brackets can be rewritten in terms of the unitary polarizations using \eq{mg_id1},
\begin{align}
\left\langle \hf^*_{\alpha\beta} (-\fvec{k}) \hf^{\alpha\beta}(\fvec{k}')\right\rangle = \sum &C_{\lp}(\omega, \omega')\\
&\times \left\langle \hf^*_{\lp}(-\fvec{k}) \hf_{\lp} (\fvec{k}') \right\rangle , \nonumber
\end{align}
for a sum over the degrees of freedom $\lp$ of \eq{h_mg_unitary}, and $C_{\lp}$ as defined in \eq{mg_id1_coeff}. 

Making use of all the usual assumptions \aref{gaussian}--\aref{isotropic} about the background, we can then use \eq{correlation_isotropic} to write $\rgw$ as
\beq
\rgw = \frac{\pi c^2}{4G} \int_0^\infty \sum_{\lp} \lambda_{\lp}(f) S_{\lp}(f)f^2 \infd f\, ,
\eeq
for $\lambda_{\lp}(f)\equiv C_{\lp}(f,f)$, and $\PSD_{\lp}(f)$ the PSDs of the unitary polarization amplitudes.
Here we have assumed that the polarization amplitudes in the unitary gauge are statistically independent, which is justified because, unlike the synchronous amplitudes, they diagonalize the kinetic matrix of the theory.
Note that we recover the GR expression, \eq{rgw_gr}, in the limit of
vanishing $\mgratio$, if we also force $S_{\lp}(f)=0$ for nontensorial
modes, which is appropriate if these additional degrees of freedom
are frozen out~\cite{DeRham2014}.

\subsection{Cross-correlation} \label{app:mg_crosscor}

Here we derive an expression, in the form of \eq{result}, for the cross-correlation of detector outputs as a function of the fractional energy spectrum of massive gravitational waves, \eq{ogw_mg}.
Going back to \eq{crosscor}, we may write the cross-correlation of the outputs of two detectors as
\begin{align}
\left\langle \hf_{\ifo}^*(f) \hf_{\ifo'}(f') \right \rangle = 
&\int \infd\dir \infd\dir' \left\langle \syn{\hf}^*_{\syn{\lp}}(\fvec{k}) \syn{\hf}_{\syn{\lp}'}(\fvec{k}') \right\rangle \\
&\times F^{*\syn{\lp}}_{\ifo} (\dir) F^{\syn{\lp}'}_{\ifo'} (\dir')\, e^{i (\vec{k}_{\syn{\lp}'}
\cdot\vec{x}_{\ifo'} - \vec{k}_{\syn{\lp}} \cdot\vec{x}_{\ifo})}\, ,\nonumber
\end{align}
where the under-tilded quantities are defined in the {\em synchronous} gauge of \eq{h_mg_synchronous_app}.
The reason we carry out the expansion in terms of the synchronous amplitudes is that only in the synchronous gauge may we write out the detector response by applying \eq{detector}.
However, we need a relation in terms of the {\em unitary} degrees of freedom, which diagonalize the kinetic matrix of the theory---we obtain such an expression below.

First, assuming a \aref{stationary} stationary and \aref{isotropic} isotropic background, with \aref{indepsky} uncorrelated sky bins, we may rewrite the above equation as (see Appendix \ref{app:correlation})
\begin{align}
\left\langle \hf_{\ifo}^*(f) \hf_{\ifo'}(f') \right \rangle = 
\delta(f-f') &\left\langle \syn{\hf}^*_{\syn{\lp}}(f) \syn{\hf}_{\syn{\lp}'}(f) \right\rangle \nonumber \\
&\times \Gamma^{\syn{\lp}\syn{\lp}'}_{~~\ifo\ifo'}(f)\, ,
\end{align}
where we have pushed all the directional dependence into the generalized overlap reduction functions, $\Gamma^{\syn{\lp}\syn{\lp}'}_{~~\ifo\ifo'}(f)$, of \eq{overlap}.
Using the transformation of \eq{h_mg_transf}, we can now write this directly in terms of the unitary polarization amplitudes,
\begin{align}
\left\langle \hf_{\ifo}^*(f) \hf_{\ifo'}(f') \right \rangle &= 
\delta(f-f') \left\langle {\hf}^*_{B}(f) {\hf}_{B'}(f) \right\rangle \\
&\times  M_{\syn{\lp}}{}^{B}(f)\, M_{\syn{\lp}'}{}^{B'}\hspace{-2pt}(f)\, \Gamma^{\syn{\lp}\syn{\lp}'}_{~~\ifo\ifo'}(f)\, . \nonumber
\end{align}
Here we have explicitly denoted the frequency dependence in
$M_{\syn{\lp}}{}^{B}(f)$, which is acquired implicitly via $\mgratio$
in \eq{h_mg_transf}.
Because the unitary polarizations can be taken to be \aref{indeppol} statistically independent, we may rewrite the above equation as a single sum over $B$,
\begin{align}
\left\langle \hf_{\ifo}^*(f) \hf_{\ifo'}(f') \right \rangle &= 
\frac{1}{2} \delta(f-f') \sum_{B} S_B(f)  \\
&\times M_{\syn{\lp}B}(f)\, M_{\syn{\lp}'B}(f)\, \Gamma^{\syn{\lp}\syn{\lp}'}_{~~\ifo\ifo'}(f)\, . \nonumber
\end{align}
Using \eq{psd_mg_unit}, this may be written directly in terms of the fractional energy spectrum for each unitary polarization as in \eq{mg_crosscor}.

Without more information about the detectors, \eq{mg_crosscor} would be our final result for massive gravity.
However, we may further simplify this for the case of a differential-arm instrument that effects a measurement via the detector tensor of \eq{detector}.
In that case, it may be shown from the definition of the antenna patterns, \eq{ap}, that $F^{\rm b}_{\ifo}(\dir) = - F^{\rm l}_{\ifo}(\dir)$ (e.g.\ \cite{Isi2017}).
This means that the generalized overlap reduction functions, \eq{overlap}, for the breathing and longitudinal modes will {\em not} be diagonal.
In fact, this is evident from our expression for the $\Gamma^{\syn{\lp}\syn{\lp}'}_{~~\ifo\ifo'}$ factors for differential-arm detectors, \eq{mg_overlap}, which follows directly from $F^{\rm b}_{\ifo}(\dir) = - F^{\rm l}_{\ifo}(\dir)$.

Using \eq{mg_overlap} and the definitions of $\lambda_B(f)$ and $M_{AB}(f)$, from \eq{lambda_mg} and \eq{h_mg_transf} respectively, our final result for the cross-correlation of the detector outputs of two differential-arm detectors takes the form of \eq{result} with $\Xi_{\lp}(f)$ implicitly defined by
\beq
\Xi_{B} (f) \thinspace \Gamma^{B}_{~\ifo\ifo'} = \lambda^{-1}_B(f) M_{\syn{\lp}B}(f) M_{\syn{\lp}'B}(f) \Gamma^{\syn{\lp}\syn{\lp}'}_{~~\ifo\ifo'}(f) .
\eeq
This reduces to the main result of \eq{xi_mg}, to quadratic order in $\mgratio$.

\bibliography{gr,gw}

\end{document}